\documentclass[preprint]{ptephy_v1}

\usepackage{amsmath} 
\usepackage{amsthm} 
\usepackage{amssymb}
\usepackage{hyperref}     
\usepackage{physics}
\usepackage{slashed,ulem,color}
\usepackage{comment}

\newcommand{\mr}[1]{\mathrm{#1}}

\graphicspath{{fig/}} 

\begin{document}

\title{A model of pseudo-Nambu-Goldstone dark matter from a softly broken $SU(2)$ global symmetry with a $U(1)$ gauge symmetry}
\author[1]{Tomohiro Abe\thanks{abe.tomohiro@rs.tus.ac.jp}}

\author[2]{Yu Hamada\thanks{yuhamada@post.kek.jp}}

\affil{
  Department of Physics, Faculty of Science and Technology, Tokyo University of Science,
  Noda, Chiba 278-8510, Japan
}

\affil[2]{
Theory Center, High Energy Accelerator Research Organization (KEK), 
Tsukuba 305-0801, Japan
}

\preprintnumber{KEK-TH-2428}

\begin{abstract}
A model of the pseudo-Nambu-Goldstone (pNG) dark matter (DM) is proposed.
We assume that there is an $SU(2)_g$ global symmetry and a $U(1)_X$ gauge symmetry in the dark sector,  
and they are spontaneously broken into a $U(1)_D$ global symmetry after a scalar field develops a vacuum expectation value.
We add a soft symmetry breaking term that breaks the $SU(2)_g$ global symmetry into the $U(1)_g$ global symmetry explicitly.
Our model predicts a stable complex pNG particle under the $U(1)_D$ global symmetry.
One of the virtues of the pNG DM models is that the models can explain the current null results in the direct detection experiments. The small momentum transfer suppresses the scattering amplitudes thanks to the low energy behavior of the Nambu-Goldstone boson. 
In our model, the soft symmetry breaking term is uniquely determined. 
This is the advantage of our model to some earlier works in which some soft symmetry breaking terms cannot be forbidden but are simply assumed to be absent to avoid the constraints from the direct detection experiments.
We calculate the thermal relic abundance of the pNG DM and find that model can explain the measured value of the DM energy density under some constraints from the perturbative unitarity. 
\end{abstract}

\maketitle


\section{Introduction}
A popular scenario about dark matter (DM) production is that 
DM particles interact with the standard model (SM) particles and were in the thermal bath made by the SM particles in the early Universe.
The interaction rate is lowered as the universe expands, and the interactions between the DM and SM particles are eventually decoupled. After the decoupling, the DM number density per comoving volume is fixed.
In this scenario, the DM in the current universe is the thermal relic, and the measured value of the DM energy density~\cite{1807.06209} requires some small but non-zero DM annihilation cross section into the SM particles. The canonical value is $\expval{\sigma v} \simeq 10^{-26}$~cm$^3$s$^{-1}$. The crossing symmetry of the Feynman diagrams implies that DM-SM elastic scatterings can also happen. 
The DM direct detection experiments have been trying to detect such scattering processes over the decades.
However, there are no clear signals of the DM-SM scatterings, and the experiments give 
stringent upper bounds on the DM-nucleon scattering cross section~\cite{1805.12562, PandaX-4T:2021bab, LUX2022}.
It is necessary to suppress the DM-SM scattering process while keeping the DM annihilation cross section if the DM particles are the thermal relic.

Pseudo-Nambu-Goldstone (pNG) DM models can explain the null results at the direct detection experiments while keeping the canonical value of the DM annihilation cross section into the SM particles~\cite{1708.02253}.
In the pNG DM models, the DM-SM scattering amplitude is proportional to $t$, where $t$ is the Mandelstam variable and is the momentum transfer squared at the low energy. 
Since the momentum transfer at the direct detection experiments is typically small, 
the scattering amplitude is highly suppressed.\footnote{However, it is not the case if the mediator mass is not larger than the momentum transfer~\cite{Abe:2021vat}.}
As a result, the models can explain the null results of the direct detection experiments.

In the model proposed in Ref.~\cite{1708.02253}, 
a Nambu-Goldstone (NG) boson is accompanied by a spontaneous symmetry breaking of a $U(1)$ global symmetry after
a complex scalar field $S$ develops a vacuum expectation value (VEV).
The global symmetry is also broken explicitly by the soft symmetry breaking term $S^2 + S^{*2}$, and thus the NG boson becomes massive. 
Other soft symmetry breaking terms are also possible but are assumed to be absent because they give unsuppressed contributions to the scattering amplitudes. 
This assumption is reasonable to explain the experimental results, but it is more satisfactory if we can forbid those terms by some symmetries.
Moreover, the soft symmetry breaking term explicitly breaks the global $U(1)$ symmetry into a global $Z_2$ symmetry, and this $Z_2$ symmetry is spontaneously broken. Hence the model~\cite{1708.02253} suffers from the domain-wall problem~\cite{Zeldovich:1974uw}.

A possible mechanism to have (almost) only the quadratic term and to avoid the domain wall problem have been already discussed in Ref.~\cite{1708.02253}. 
Models that explain why only the $S^2 + S^{*2}$ term arises are proposed in Refs.~\cite{2001.03954,2001.05910, 2104.13523, 2105.03419}.
Those models utilize the $U(1)_{B-L}$ gauge symmetry to restrict the soft symmetry breaking terms. 
The $U(1)_{B-L}$ gauge symmetry is spontaneously broken at high energy, and 
the pNG model proposed in Ref~\cite{1708.02253} arises as the low energy effective theory.\footnote{Note that, in general, the domain-wall problem can be avoided if the soft breaking term arises from a VEV that spontaneously breaks a continuous (simple) symmetry $G$ into the $Z_2$ symmetry,
in which $G$ is not needed to be gauged.
In fact, suppose that two scalar VEVs trigger a two-step symmetry breaking $G \to Z_2 \to 1$, 
in which the pNG model in Ref~\cite{1708.02253} arises after the first symmetry breaking.
Since $\pi_0(G)=0$ and $\pi_1(G/Z_2)=Z_2$,
cosmic strings with the topological charge $Z_2$ appear at the first phase transition and become connected by domain walls at the second one.
Regardless of whether $G$ is global or local, the domain walls shrink well before they dominate the energy density of the Universe, see Sec.~13.6 in Ref.~\cite{Vilenkin:2000jqa} and also Refs.~\cite{Everett:1982nm,Kibble:1982dd}.
}
The models predict that the DM is not absolutely stable and decays eventually. 
To make the DM long-lived particle, the scale of the $U(1)_{B-L}$ breaking is assumed to be very high around the GUT scale.
In Ref.~\cite{2101.04887}, the effect of the mass-dimension one soft-breaking term, $S + S^*$ is discussed. 
In that case, the $Z_2$ symmetry is also explicitly broken, and thus the domain-wall problem is absent. 
However, the degeneracy in the mass spectra of the scalar particles is necessary to suppress the DM-quark scattering amplitude. Analyses with other explicit breaking terms are discussed in Refs.~\cite{Alanne:2020jwx, 2106.05289}.

In this paper, we construct a model that allows only the mass-dimension-two soft symmetry breaking terms by a gauge symmetry. The gauge symmetry is broken around the electroweak or TeV scale in contrast to the pNG models with the $U(1)_{B-L}$ gauge symmetry. In our model, any discrete symmetry is not spontaneously broken and the domain-wall problem is absent.

The rest of this paper is organized as follows.
In Sec.~\ref{sec:model}, we introduce our model setup.
In Sec.~\ref{sec:scattering-amplitude}, we show that the DM-SM elastic scattering amplitude is suppressed at the tree level.
Some constraints from the perturbative unitarity, Higgs invisible decay, and cosmic strings are discussed in Sec.~\ref{sec:constraints}.
We show that the model can explain the measured value of the DM energy density in Sec.~\ref{sec:relic}.
In Section~\ref{sec:loop}, we consider the direct detection of the pNG DM at the loop level.
In Section~\ref{sec:two-component}, we also consider a two-component DM scenario in our model.
Section~\ref{sec:summary} is devoted for the conclusion.
We present cosmic string solutions in our model in Appendix.~\ref{sec:appendix}.
The calculated results of the scattering amplitudes in the direct detection at the loop level are provided in Appendix.~\ref{sec:appendix2}.

\section{Model}\label{sec:model}
We introduce a complex scalar field $\phi$ that transforms under an $SU(2)_g$ global symmetry and a $U(1)_X$ gauge symmetry as\footnote{A model utilizing an $SU(2)$ global symmetry without any gauge symmetries is discussed in Ref.~\cite{Karamitros:2019ewv}.}
\begin{align}
\phi \to e^{i T^a \theta_g^a} e^{i \frac{1}{2} \theta_X(x)} \phi,
\end{align}
 where $T^a$ is the $SU(2)_g$ generator.
Under the symmetry of the SM sector, $\phi$ is a singlet.
All the SM particles are singlets under the $SU(2)$$_g \times$ $U(1)_X$ symmetry.
We assume that $\phi$ develops a VEV,
 \begin{align}
\expval{\phi} = \begin{pmatrix} 0 \\ \frac{v_s}{\sqrt{2}} \end{pmatrix}, 
 \end{align}
 and 
 breaks the symmetry spontaneously as $SU(2)_g \times U(1)_X \to U(1)_D$,
 where $U(1)_D$ is a global symmetry and is expressed as
\begin{align}
\phi \to e^{i T^3 \theta_D} e^{i \frac{1}{2} \theta_D} \phi
=
\begin{pmatrix}
e^{i \theta_D} & 0 \\ 0 & 1 
\end{pmatrix}
 \phi.
\end{align}
This spontaneous symmetry breaking generates three NG bosons. 
One is the would-be NG boson that is eaten by the $U(1)_X$ gauge boson.
The other two are massless NG bosons.

To make the two NG bosons massive, 
we introduce an explicit symmetry breaking term that softly breaks $SU(2)_g$ into $U(1)_g$, 
\begin{align}
 \phi^\dagger T^3 \phi.
 \label{eq:soft-breaking} 
\end{align}
This soft-breaking term\footnote{
We can consider $\sum_a c^a \phi^\dagger T^a \phi$, which looks more general soft breaking terms.
However, after the field redefinition, $\phi \to e^{i T^a \theta^a} \phi$,
these terms transform as
$
 \sum_a c^a \phi^\dagger T^a \phi 
\to
\sum_{a,b}
O^{ba} c^a 
\phi^\dagger 
T^b
 \phi 
,
$
where
$O^{ba} = \left( e^{-i X^c \theta^c}\right) _{ba}$, and $(X^c)_{ab} = - i\epsilon^{cab}$.
Since $O$ is a three-by-three orthogonal matrix, the norms of $\vec{c}$ and $O \vec{c}$ are the same. Thus we can make $O \vec{c} = (0,0, \abs{\vec{c}})$ by choosing $\vec{\theta}$ appropriately. 
Therefore, we can take $c^a = c \delta^{a3}$ without loss of generality
and obtain Eq.~\eqref{eq:soft-breaking} 
}
 changes the symmetry breaking pattern.
After $\phi$ develops the VEV,  $U(1)$$_g\times$ $U(1)_X$ is spontaneously broken into $U(1)_D$,
and we obtain a would-be NG boson and two pNG bosons.
The $U(1)_D$ global symmetry makes the pNG bosons stable,
and thus these two pNG bosons are DM candidates.
Thanks to the $U(1)_X$ gauge symmetry, we cannot write down other soft symmetry breaking terms that explicitly break the global $SU(2)_g$ symmetry.
Therefore, we can explain why the soft symmetry breaking term is the mass-dimension-two in our setup.

We also impose that the new sector is symmetric under the charge conjugation,
\begin{align}
 \phi \to& \phi^*, \label{eq:dark-charge-conjugation-1}\\
 V_\mu \to& - V_\mu, \label{eq:dark-charge-conjugation-2}
\end{align}
where $V_\mu$ is the $U(1)_X$ gauge boson.
All the SM particles are singlets under this charge conjugation. 
This symmetry is $Z_2$ but not broken by the VEV of $\phi$.
This symmetry also forbids the gauge kinetic mixing between the $U(1)_X$ and $U(1)$$_Y$ gauge fields.\footnote{If we do not impose the $Z_2$ symmetry and introduce the gauge kinetic mixing, then the amplitude of the pNG DM scattering of the SM particle is unsuppressed by the small momentum transfer~\cite{2109.11499}
}

Now we have two DM candidates. One is the pNG DM that is stabilized by the $U(1)_D$ global symmetry and the $Z_2$ symmetry.
The other is the gauge boson $V_\mu$ that can be stabilized by the $Z_2$ symmetry.
The model contains a multi-component DM scenario depending on the mass spectra.
If the mass of $V$ is larger than twice the mass of the pNG DM, namely
$m_V > 2 m_\chi$, then $V$ can decay into $\chi \chi^\dagger$ and thus only the pNG bosons are the DM candidates.
On the other hand, if $m_V < 2 m_\chi$, then $V$ cannot decay due to the $Z_2$ symmetry and $\chi$ cannot decay due to the $U(1)_D$ symmetry, and thus the model has two DM components.

\subsection{Lagrangian}

Under the setup, the renormalizable Lagrangian is given by
\begin{align}
 {\cal L}
=&
 \left. {\cal L} \right|_{\text{SM w/o Higgs potential}}
+ D^\mu \phi^\dagger D_\mu \phi
- \frac{1}{4} V^{\mu \nu} V_{\mu \nu}
-V_{\text{SU(2)-global}}
-V_{\text{soft}}
,
\label{eq:lagrangian}
\end{align}
where
\begin{align}
 D_{\mu} \phi
=& \partial_\mu \phi
 + i \frac{1}{2} g_D V_\mu \phi
,\\
 V_{\mu \nu}
=& \partial_\mu V_\nu - \partial_\nu V_\mu
,\\
V_{\text{SU(2)-global}}
=&
+ \mu_H^2 H^\dagger H  
+ \mu_\phi^2 \phi^\dagger \phi
\nonumber\\
&
+ \lambda_H \left( H^\dagger H \right)^2
+ \lambda_\phi \left(\phi^\dagger \phi \right)^2
+ \lambda_{H\phi} ( H^\dagger H ) \left(\phi^\dagger \phi \right)
,\\
V_{\text{soft}}
=&
\mu_\chi^2 \left( \phi^\dagger T^3 \phi \right)
, \label{eq:soft-term}
\end{align}
and $H$ is the SM Higgs field.
As we have already discussed, the soft breaking term, $V_{\text{soft}}$, contains only a mass-dimension-two operator. Mass-dimension-one operators and mass-dimension-three operators are forbidden by the $U(1)_X$ gauge symmetry.

We make a brief comment on mass-dimension-four operators that explicitly break the $SU(2)_g$ global symmetry.
In our setup, we neglect those terms to explain the current null results at the dark matter direct detection experiments.
Suppose the global $SU(2)_g$ is exact. Then, the NG boson scattering amplitude is suppressed by the small momentum at the low energy. This is also true if the global $SU(2)_g$ is broken by the mass dimension-2 operator because it does not modify the structure of the interaction. We utilize this nature in our model to suppress the DM-quark scattering amplitude, which we will show in Sec.~\ref{sec:scattering-amplitude}.
If we add the mass-dimension-four global $SU(2)_g$ breaking terms, the structure of the interactions is modified, and the scattering amplitude is not suppressed at low energy in general as discussed in Ref.~\cite{1708.02253}.

\subsection{scalar masses and couplings}

Component fields and VEVs are parametrized as
\begin{align}
 H = \begin{pmatrix}
      i \pi_{W^+} \\  \frac{v + \sigma - i \pi_Z}{\sqrt{2}}
     \end{pmatrix}
, \ 
 \phi = 
\begin{pmatrix}
i \chi \\    
\frac{1}{\sqrt{2}} \left( v_s + s - i \chi_{V} \right)
     \end{pmatrix}
,
\end{align}
where $v$ and $v_s$ are the VEVs,  
$\pi_{W^+}$, $\pi_Z$, and $\chi_V$ are the would-be NG bosons for $W^+$, $Z$, and $V$, respectively.
The stationary condition of this vacuum imposes the following relations for the mass parameters,
\begin{align}
 \mu_H^2 =& -v^2 \lambda_H - \frac{1}{2} v_s^2 \lambda_{H\phi} , \\ 
 \mu_\phi^2 =& \frac{\mu_\chi^2}{2} - \frac{1}{2} v^2 \lambda_{H\phi}  - v_s^2 \lambda_\phi. 
\end{align}
The mass terms of the physical scalar particles are given by
\begin{align}
 {\cal L}_\text{mass}^\text{scalar}
=&
- \mu_\chi^{2} \chi^\dagger \chi
- \frac{1}{2} 
 \begin{pmatrix}  \sigma & s \end{pmatrix}
 \begin{pmatrix}
  2 \lambda_H v^2  &  \lambda_{H\phi} v v_s \\
  \lambda_{H\phi} v v_s  &  2 \lambda_\phi v_s^2 
 \end{pmatrix}
 \begin{pmatrix}  \sigma \\  s \end{pmatrix}
.
\end{align}
It is found that the mass of the DM $m_\chi$ is given by $\mu_\chi$, namely $m_\chi^2 = \mu_\chi^2$.
The mass eigenstates that are denoted by $h$ and $h'$ are obtained by diagonalizing the
two-by-two mass matrix above.
The relation between the mass eigenstates and component fields are given by 
\begin{align}
  \begin{pmatrix}  \sigma \\ s  \end{pmatrix}
=&
\begin{pmatrix}
  c_\theta  & s_\theta \\
 -s_\theta  & c_\theta 
\end{pmatrix}
 \begin{pmatrix}  h \\ h' \end{pmatrix}
,
\end{align}
where
 $c_{\theta} = \cos\theta_h$
and 
 $s_{\theta} = \sin\theta_h$.
The mass eigenvalues and the mass matrix is related by the mixing angle as follows.
\begin{align}
\begin{pmatrix}
  2 \lambda_H v^2  &  \lambda_{H\phi} v v_s \\
  \lambda_{H\phi} v v_s  &  2 \lambda_\phi v_s^2 
\end{pmatrix}
=
\begin{pmatrix}
  c_\theta  & s_\theta \\
 -s_\theta  & c_\theta 
\end{pmatrix}
\begin{pmatrix}
  m_h^2  & 0 \\
  0  & m_{h'}^2
\end{pmatrix}
\begin{pmatrix}
  c_\theta  & -s_\theta \\
 s_\theta  & c_\theta 
\end{pmatrix}
,
\end{align}
where $m_h$ and $m_{h'}$ are the masses of $h$ and $h'$, respectively.
The quartic couplings are expressed by $m_h$, $m_{h'}$, and $\theta_h$ as
\begin{align}
 \lambda_H =& \frac{m_h^2 c_\theta^2 + m_{h'}^2 s_\theta^2}{2 v^2},\\
 \lambda_\phi =& \frac{m_h^2 s_\theta^2 + m_{h'}^2 c_\theta^2}{2 v_s^2},\\
 \lambda_{H\phi} =& \frac{(m_{h'}^2 - m_h^2) s_\theta c_\theta}{v v_s}.
\end{align}

The $\chi$-SM fermion elastic scattering processes are mediated by exchanging $h$ and $h'$.
The relevant couplings for the processes are given by
\begin{align}
 {\cal L}\supset&
 - g_{\chi \chi h} \chi^\dagger \chi h
 - g_{\chi \chi h'} \chi^\dagger \chi h'
 - g_{\bar{f} f h} \bar{f} f h 
 - g_{\bar{f} f h'} \bar{f} f h' 
,
\end{align}
where
\begin{align}
 g_{\chi \chi h}
=& - \frac{m_h^2}{v_s} s_\theta
\label{eq:g_chichih}
,\\
 g_{\chi \chi h'}
=& + \frac{m_{h'}^2}{v_s} c_\theta
 ,\\
 g_{\bar{f} f h}
=& + \frac{m_{f}}{v} c_\theta
 ,\\
 g_{\bar{f} f h'}
=& + \frac{m_{f}}{v} s_\theta.
\end{align}
Here $f$ stands for the SM fermions.

\subsection{gauge boson mass and interaction terms}
After $\phi$ develops the VEV, the $U(1)_X$ gauge boson acquires mass,
\begin{align}
 {\cal L}
\supset&
\frac{m_V^2}{2} V^\mu V_\mu
,
\end{align}
where
\begin{align}
 m_V^2 =& \frac{1}{4} g_D^2 v_s^2.
\end{align}

The relevant interaction terms containing $V$ for the following analysis are given by 
\begin{align}
{\cal L} 
\supset& 
- i g_{V\chi\chi} V^\mu \left( \chi^\dagger \overset{\leftrightarrow}{\partial}_\mu \chi \right)
+ \frac{1}{2} g_{VVh} V^\mu V_\mu h
+ \frac{1}{2} g_{VVh'} V^\mu V_\mu h'
,
\end{align}
where
\begin{align}
g_{V\chi\chi}
=&
 \frac{m_V}{v_s}
,
\\
g_{VVh} 
=& - \frac{2 m_V^2}{v_s} s_\theta,
\\
g_{VVh'} 
=&  \frac{2 m_V^2}{v_s} c_\theta.
\end{align}

\subsection{Parameters}

There are seven model parameters in the scalar and the new gauge sectors,
\begin{align}
\left(
\mu_H^2,\ \mu_s^2,\   \lambda_H,\ \lambda_\Phi,\ \lambda_{H\Phi},\ \mu_\chi^2,\ g_D
\right).
\end{align}
Instead of using these parameters, we choose the following seven parameters as inputs in the following analysis.
\begin{align}
\left( v,\ v_s,\   m_h,\ m_{h'},\ \theta_{h},\ m_\chi,\ m_{V} \right),
\end{align}
Among these parameters, $v$ and $m_h$ are already known, $v \simeq 246$~GeV, and $m_h \simeq$ 125~GeV~\cite{Zyla:2020zbs}, 
and thus we have five free parameters 
$\left( \ v_s,\  m_{h'},\ \theta_{h},\ m_\chi,\ m_{V} \right)$.

\section{Tree-level DM-quark scattering amplitudes}\label{sec:scattering-amplitude}

The DM scattering off the quark is the essential process for the DM direct detection experiments.
This $\chi q$ scattering process is mediated by $h$ and $h'$ exchanging diagrams in the $t$-channel.
Its amplitude is given by
\begin{align}
i {\cal M}
=& 
-ig _{\chi \chi h} \frac{i}{t - m_h^2} (-i g_{\bar{f} fh}) \bar{u} u
-ig _{\chi \chi h'} \frac{i}{t - m_{h'}^2} (-i g_{\bar{f} fh'}) \bar{u} u
\nonumber\\
=& 
-i
\frac{m_f}{v v_s} s_\theta c_\theta
\left(
- \frac{m_h^2}{t - m_h^2}  + \frac{m_{h'}^2}{t - m_{h'}^2} 
\right)
\bar{u} u
\nonumber\\
=& 
-i
\frac{m_f}{v v_s} s_\theta c_\theta 
\left(
- \frac{t}{t - m_h^2}  + \frac{t}{t - m_{h'}^2} 
\right)
\bar{u} u
.
\end{align}
Since $\abs{t} \ll m_{h}^2, m_{h'}^2$ at the direct detection experiments, this amplitude is suppressed by $\frac{t}{m_h^2}$ and $\frac{t}{m_{h'}^2}$.
As a result, the model predict very small spin-independent cross section at the direct detection experiments.
This is the virtue of the pNG DM models.

At the loop level, however, the scattering amplitude that is not suppressed by the small momentum transfer can be induced~\cite{Ishiwata:2018sdi, Azevedo:2018exj, Glaus:2020ihj}.
We discuss the loop effect in Sec.~\ref{sec:loop}

For $m_V < 2 m_{\chi}$, both $\chi$ and $V$ are stable. 
Then $V$ is also a DM component, and we have to consider the $Vq$ scattering process.
The scattering amplitude is given by 
\begin{align}
i {\cal M}
=& 
ig _{VV h} \frac{i}{t - m_h^2} (-i g_{\bar{f} fh}) \bar{u} u
+ ig _{VV h'} \frac{i}{t - m_{h'}^2} (-i g_{\bar{f} fh'}) \bar{u} u
\nonumber\\
=& 
i
\frac{2 m_f}{v v_s} s_\theta c_\theta
\left(
- \frac{m_V^2}{t - m_h^2}  + \frac{m_V^2}{t - m_{h'}^2} 
\right)
\bar{u} u
.
\end{align}
In contract to $\chi q$ scattering, the $Vq$ scattering amplitude is not suppressed.
Therefore, the results from the DM direct detection experiments give constraints on the mass and the number density of $V$. 

The purpose of the current paper is to investigate whether the model given in Eq.~\eqref{eq:lagrangian}
is a viable pNG DM model. Thus we mainly focus on the single component DM scenario where $m_V > 2 m_\chi$.
For the multi-component scenario where $m_V < 2 m_\chi$ is briefly discussed in Sec.~\ref{sec:two-component}

\section{Constraints}\label{sec:constraints}

In this section, we discuss constraints from the perturbative unitarity, the Higgs invisible decay, and cosmic strings.

\subsection{Perturbative unitarity}

We find the constraints on the scalar and gauge couplings from the perturbative unitarity (PU) bound~\cite{Lee:1977eg}.
Two-to-two scattering amplitudes at high energy can be decomposed into partial waves,
\begin{align}
 {\cal M}
= 16 \pi \sum_\ell (2 \ell + 1) P_\ell(\cos\theta) a_\ell,
\end{align} 
where $P_\ell$ is the Legendre polynomial, and $\cos \theta$ is the scattering angle.
In general, there are inelastic scatterings, and thus $a_\ell$ is a matrix.
Unitarity requires each eigenvalue of $a_\ell$ to satisfy
\begin{align}
 \abs{\Re a_\ell} \leq \frac{1}{2}.
\end{align}
We utilize this inequality to find the constraints on the scalar and gauge couplings.

\subsubsection{scalar quartic couplings}
Two-to-two scattering processes containing only the scalar bosons give the following PU bound on the scalar quartic couplings. 
\begin{align}
 |\lambda_H| <& 4\pi,\\
 |\lambda_{H\phi}| <& 8\pi,\\
 |\lambda_\phi| <& 4\pi,\\
 \abs{3 \lambda_H + 3\lambda_\phi \pm \sqrt{9(\lambda_\phi - \lambda_H)^2 + 4 \lambda_{H \phi}} } <& 8\pi.\label{eq:PU-scalar-4}
\end{align}
In our following analysis, $\lambda_\phi$ is typically the largest quartic coupling among the three quartic couplings. 
In that case, the condition given in Eq.~\eqref{eq:PU-scalar-4} is essential and is reduced to
\begin{align}
 \abs{\lambda_\phi} < \frac{4\pi}{3}. 
\label{eq:PU-lambda_phi}
\end{align}

\subsubsection{gauge coupling}
We study $V \chi \to V \chi$ scattering at high energy in order to find the constraint on $g_D$. Here $V$ is transversely polarized. The longitudinal mode is already taken into account in the previous subsection thanks to the equivalence theorem~\cite{Cornwall:1974km, Lee:1977eg, Chanowitz:1985hj,Gounaris:1986cr,Veltman:1989ud}.
We find
\begin{align}
 a_0^{\lambda \lambda'} =& - \frac{g_D^2}{96\pi} \lambda \lambda',\\
 a_1^{\lambda \lambda'} =& - \frac{g_D^2}{192\pi} \lambda \lambda',\\
 a_2^{\lambda \lambda'} =& - \frac{g_D^2}{960\pi} \lambda \lambda',
\end{align}
where $\lambda$ and $\lambda'$ are the helicity of $V$ in the initial and the final states, respectively. 
The $s$-wave ($\ell = 0$) gives the stringent bound.
Since there are two transversely polarized states, $a_0$ is a two-by-two matrix.
Its eigenvalues are $0$ and $-\frac{g_D^2}{48\pi}$, and their absolute values should be smaller than $1/2$. We find
\begin{align}
 g_D < \sqrt{24 \pi}. \label{eq:PU-gauge}
\end{align}

\subsection{Higgs invisible decay}
If $m_\chi < m_h/2$, then the Higgs boson decay into $\chi \chi^\dagger$.
The decay width is given by
\begin{align}
 \Gamma(h \to \chi \chi^\dagger)
=&
 \frac{1}{16 \pi} \frac{g_{\chi\chi h}^2}{m_h} \sqrt{1 - \frac{4 m_\chi^2}{m_{h}^2}}
 \theta(m_{h} - 2 m_\chi)
.
\end{align}
This process is the Higgs invisible decay 
and is being searched by the ATLAS and CMS experiments. 
Currently, the ATLAS and CMS experiments obtain the upper bound on it as 
\begin{align}
 \text{BR}_\text{inv} <  
\begin{cases}
 0.11 & \text{(ATLAS \cite{ATLAS-CONF-2020-052})} \\
 0.18 & \text{(CMS \cite{CMS:2022qva})} 
\end{cases}
\end{align}
at 95\% CL.
The prospects of various experiments are summarized in \cite{1905.03764},
\begin{align}
 \text{BR}_\text{inv} <
\begin{cases}
 0.019 & \text{(HL-LHC)} \\
 0.0026 & \text{(ILC(250))} \\
 0.00024 & \text{(FCC)}
\end{cases}
\end{align}
at 95\% CL,
where FCC corresponds to the combined performance of 
FCC-ee$_\text{240}$,
FCC-ee$_\text{365}$,
FCC-eh, and FCC-hh.
The prospects for the ILC and FCC are obtained by combining with the HL-LHC.

\subsection{DM relic abundance from cosmic strings}
Our model admits cosmic strings (or vortex strings)~\cite{Vilenkin:2000jqa} as non-trivial solutions of the static equations of motion.
In fact, concentrating on the dark sector, the symmetry breaking pattern $U(1)_g\times U(1)_X \to U(1)_D$ leads to the non-trivial first homotopy group of the vacuum
\begin{align}
    \pi_1 \left(\frac{U(1)_g\times U(1)_X}{U(1)_D} \right) \simeq \pi_1 \left(U(1) \right) = \mathbb{Z} \, ,
\end{align}
which means that the solutions are characterized by the winding number $n\in \mathbb{Z}$.
This model is similar to a model in Ref.~\cite{Eto:2016mqc} that admits semilocal strings~\cite{Vachaspati:1991dz,Achucarro:1999it} with an explicit $SU(2)$-breaking term, which is a mass-dimension-four operator like $\left(\phi^\dagger T^3 \phi\right)^2$.
In our model, however, the $SU(2)_g$ symmetry is explicitly broken by the soft breaking term, and hence the property of the string solution is different from those.
We present the string solution in our model in Appendix \ref{sec:appendix}.
In this section, we discuss their effects on cosmology,
in which the detailed structure or properties of the string solution are irrelevant.

When the temperature $T$ of the thermal bath goes below $v_s$, the $U(1)_g\times U(1)_X$ symmetry breaks spontaneously into $U(1)_D$, leading to creation of the cosmic strings~\cite{Kibble:1976sj,Zurek:1985qw}.
The created strings form a network, whose typical length scale becomes the Hubble length $l_H(T\simeq v_s)$ within the Hubble time $t_H(T\simeq v_s)$ after creation.
These strings randomly move in the expanding Universe and produce closed loops by their reconnection.
Thanks to the production of the loops, the number of the strings per Hubble patch always remains to be $\mathcal{O}(1)$, which is known as the scaling solution~\cite{Vilenkin:2000jqa}, avoiding the overclosure of the Universe.
Therefore, the string network itself does not affect the cosmological history.

On the other hand, the produced string loops could affect the DM relic abundance.
This is because the string loops decay into light particles or gravitational waves, the former of which is nothing but the pNG DM $\chi$.
Thus the string loops give an additional contribution to the DM relic abundance besides the thermal relic one.
We give a simple estimation for the former one.\footnote{
Note that, since our string is the local one and the pNG boson is associated with the global $SU(2)_g$ symmetry,
we cannot apply straightforwardly the studies on NG bosons from global $U(1)$ strings~\cite{Vilenkin:1986ku,Battye:1993jv}.
}

The time evolution of the string network can be understood analytically by the so-called velocity-dependent one-scale model~\cite{Martins:1996jp,Martins:2000cs}.
At cosmological time $t$, closed loops are produced from the network with a distribution function~\cite{Martins:2000cs}
\begin{align}
    f_\mr{loop}(l,t) = \frac{A}{\alpha t^4} \delta(l-\alpha t) \label{eq:VOSloop}
\end{align}
such that the number density of the loops produced per time with length in the range between $l$ and $l+\mathrm{d} l$ is given by $f_\mr{loop}(l,t) \mathrm{d}l$.
Here $A$ and $\alpha$ are $\mathcal{O}(0.1)$ constants.

We consider the most ``pessimistic'' case, i.e., all of such loops decay by emitting only the pNG boson,
which means that all energy of the loops is converted into the DM relic abundance.
For simplicity, we further assume that the closed loops decay instantaneously.
Note that, before the interaction between the DM and the thermal bath is out of equilibrium, $t<t_\mr{f.o.}$, the DM supplied from the loops is washed out and its number density does not deviate from the thermal equilibrium value.
Thus it is sufficient to consider only stages after the decoupling $t\geq t_\mr{f.o.}$.
Then we can find that the DM number density produced from the string loops, $n_\mr{DM, st}$, follows the Boltzman equation
\begin{align}
 \frac{\mr{d}}{\mr{d} t}n_\mr{DM, st} (t) + 3 H(t)\, n_\mr{DM, st} (t) &= \int \mr{d} l \, \frac{l \mu f_\mr{loop}(l,t)}{m_\chi}, \label{eq:DMBoltzman} 
\end{align}
where $\mu$ is the tension of the strings $\mu \sim v_s^2$ and $H(t)$ is the Hubble parameter. 
Substituting Eq.~\eqref{eq:VOSloop} and introducing the yield $Y_\mr{DM,st}\equiv n_\mr{DM, st}/s$ with $s$ the entropy density $s\sim g_\ast T^3$,
Eq.~\eqref{eq:DMBoltzman} is rewritten as
\begin{align}
 \frac{\mr{d}}{\mr{d} t} Y_\mr{DM,st} (t) 
& \simeq \frac{ \mu A}{m_\chi t^3 g_\ast T^3}  , \label{eq:DMBoltzman2} 
\end{align}
leading to
\begin{align}
Y_\mr{DM,st} (\infty) & \simeq \int_{t_\mr{f.o.}}^\infty \mr{d} t \, 
\frac{ \mu A}{m_\chi t^3 g_\ast T^3} \\
 &\sim \frac{T_\mr{f.o.}}{m_\chi} \left(\frac{v_s}{M_\mr{pl}} \right)^2 ,
\end{align}
where we have used $T^2\sim M_\mr{pl.}/t$ and $T_\mr{f.o.}$ is the freeze-out temperature known to be given as $T_\mr{f.o.}\simeq m_\chi/25$.
Therefore, compared to the correct thermal relic abundance, $ Y_\mr{DM}\sim 10^{-12}$, 
the abundance from the loops is tiny and hence does not affect the argument presented in the next section.

\section{Relic abundance}\label{sec:relic}
We show that the model explains the DM energy density by the thermal relic abundance of $\chi$ and $\chi^\dagger$.
A $\chi \chi^\dagger$ pair annihilates into the SM particles mainly by exchanging $h$ and $h'$ in the $s$-channel.
It also annihilates into $hh'$ and $h'h'$ pairs if these processes are kinematically allowed.
We calculate the relic abundance for a given parameter set and find the value of $v_s$ that produces the 
right amount of the DM as the thermal relic. 
As we mentioned at the last in Sec.~\ref{sec:scattering-amplitude}, 
we focus on the parameter space where $m_V > 2 m_\chi$.
In the following analysis, we set $m_V = 3 m_\chi$.

\begin{figure}
\centering
\includegraphics[width=0.5\hsize]{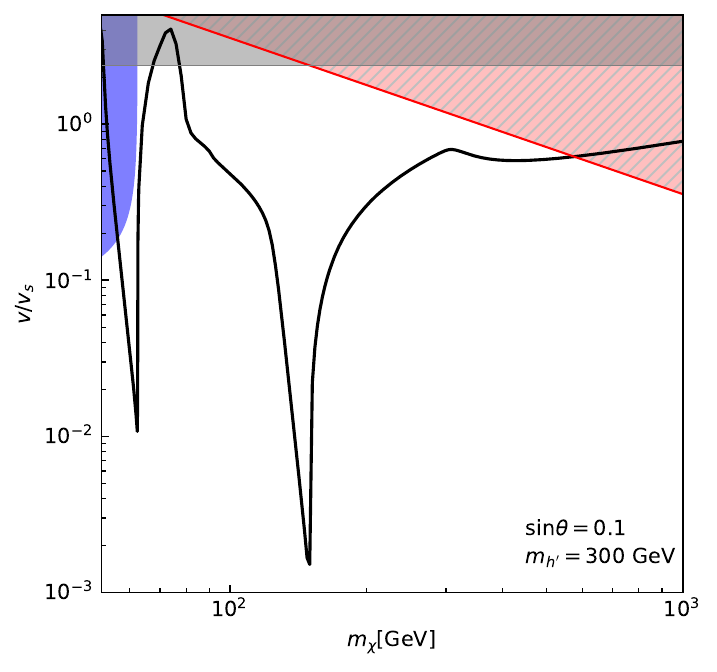}
\caption{
The black solid curve shows the value of $v_s$ that can explain the measured value of the dark matter energy density. In the gray shaded region, the perturbative unitarity for the scalar channel is violated.
In the red hatched-shaded region, the gauge coupling exceeds the value required from the perturbative unitarity.
The blue shaded region around the top-left corner is excluded by the Higgs invisible decay search.
}
\label{fig:omegah2}
\end{figure}

Figure \ref{fig:omegah2} shows the value of $v_s$ that can explain the DM energy density by the thermal relic. We use \texttt{micrOMEGAs}~\cite{Belanger:2018ccd} to calculate the relic abundance.
The constraints from the PU bound and the Higgs invisible decay are also shown. 
The PU bound for the scalar quartic couplings excludes the small $v_s$ region.
This bound is also sensitive to the choice of the mixing angle $\theta$. 
A smaller mixing angle requires smaller $v_s$ to obtain the right amount of the thermal relic.
Consequently, a too small mixing angle is disfavored by the PU bound for the scalar couplings.
The PU bound for the gauge coupling is essential for the heavy $\chi$ regime. 
It gives the upper bound on $m_\chi$.
There are two dips in Fig.~\ref{fig:omegah2}. One is at $m_\chi \simeq m_h/2$, and the other is at $m_\chi \simeq m_{h'}/2$. 
Pairs of the DM particles annihilate by exchanging $h$ and $h'$ in the $s$-channel and the annihilation cross section is enhanced by the $h$ or $h'$ resonance at those mass range. It is required to make the DM-scalar coupling smaller to obtain the right amount of the DM energy density at the resonant regions. This is the reason why we see the two dips in Fig.~\ref{fig:omegah2}. 

We focus on the Higgs funnel region where $m_\chi$ is slightly smaller than half of the SM-like Higgs boson mass. In this region, the decay of the Higgs boson into two DM particles is allowed. This Higgs invisible decay has been searched at the ATLAS and CMS experiments and will be searched at the future collider experiments as well. Hence it is worth having a closer look.

In the Higgs funnel region, the DM particles in the early universe can annihilate into the SM particles efficiently because of the Higgs resonance. The resonant enhancement requires a smaller coupling of the DM to the Higgs boson to obtain the right amount of the DM thermal relic. That is the reason why we can see the sharp dip at $m_\chi \simeq m_h/2.$ 
This small coupling makes the DM-SM elastic scattering cross section small in the early universe.
As a result, the kinetic decoupling may happen earlier than usual~\cite{1706.07433}.
One of the authors showed that the effect of the kinetic decoupling is sizable in the Higgs resonant regime in the simple pNG DM model~\cite{2106.01956}. 
We use the same method to estimate the value of $v_s$ that reproduces the measured value of the DM relic abundance. For the technical details, see Ref.~\cite{2106.01956}.\footnote{In Ref.~\cite{2106.01956},
the distribution function of the DM particle is assumed to be $f(T,E) = \alpha(T_\chi) e^{-E/T_\chi}$, where $T_\chi$ is the DM temperature. This assumption would be justified if the DM self-scattering is large enough. The recent discussion on the effect of DM self-scattering is discussed in Ref.~\cite{2204.07078}.
The method without introducing the temperature of DM, see \cite{Ala-Mattinen:2019mpa, Ala-Mattinen:2022nuj}.
}
 
\begin{figure}[htbp]
\centering
\includegraphics[width=0.5\hsize]{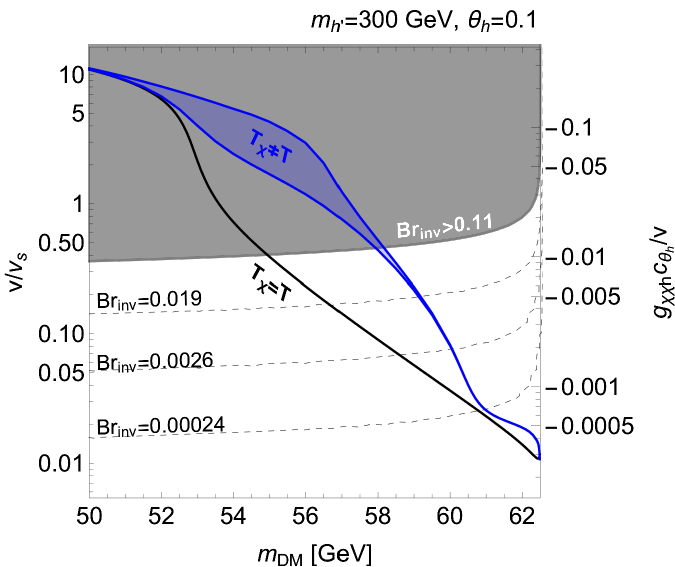}
\caption{
The value of $v_s$ that can explain the measured value of the dark matter energy density for a given DM mass in the Higgs funnel region. 
The black solid curve is obtained by the standard calculation~\cite{Gondolo:1990dk}.
The blue region is obtained by taking into account the early kinetic decoupling effect.
The lower boundary of the blue region is determined with the assumption that all the quarks ($u,d,s,c,b$) contribute to the DM elastic scattering process, while the upper boundary is determined with the assumption that only the light quarks ($u,d,s$) contribute to the scattering process~\cite{1706.07433}.
The gray shaded region is excluded by the Higgs invisible decay search by the ATLAS experiment~\cite{ATLAS-CONF-2020-052}.
The dashed curves show the prospects provided in Ref.~\cite{1905.03764}.
}
\label{fig:funnel}
\end{figure}

Figure~\ref{fig:funnel} shows the result in the Higgs funnel region.
The black solid curve is the result obtained in the standard calculation with the assumption that the kinetic equilibrium is maintained during the freeze-out.
The blue curves are obtained without assuming the kinetic equilibrium.
As the same as the original pNG DM model, we find that the effect of the kinetic decoupling is significant and the coupling may be underestimated if we naively assume the kinetic equilibrium in the calculation for the thermal relic. We also show the current bound and future prospects for the Higgs invisible decay search. We find that the decay branching ratio of the Higgs invisible decay can be enhanced more than an order of magnitude.

\section{Direct detection at loop level}\label{sec:loop}
As stated in Sec~\ref{sec:scattering-amplitude},
the scattering amplitude between the pNG DM $\chi$ and  quarks are suppressed due to the cancellation of the $h,h'$ exchanges at the tree level.
At the loop-level, on the other hand, $\chi$ has a non-zero scattering amplitude with nucleons.
We here calculate the loop-level amplitude and the spin-independent (SI) cross section with nucleons, and compare the result with the current experimental bound.

The effective Lagrangian relevant for the scattering process is given as
\begin{equation}
 \mathcal{L}_{eff} = C_q^S \, m_q \chi^\dagger \chi \bar q q + C_g^S \, \frac{\alpha_s}{\pi} \chi^\dagger \chi G^a_{\mu\nu} G^{a \mu\nu},
 \end{equation}
where $C_q^S $ and $C_g^S $ are Wilson coefficients.
We consider only the scalar-type interaction and have ignored the twist-type one since the latter is negligible.

Then the SI cross section is given as
\begin{equation}
    \sigma_\mathrm{SI} = \frac{1}{4\pi} \left( \frac{\mu_N m_N}{m_\chi}\right)^2 
\left| \sum_{q=u,d,s} C_q^S f^N_q - \frac{8}{9} C_g^S f_g^N\right|^2 ,
\end{equation}
where $m_N$ and $\mu_N\equiv m_N m_\chi/(m_N + m_\chi)$ are the nucleon mass and the reduced mass, respectively.
$f^N_q$ and $f_g^N$ are the matrix elements of the operators evaluated by the nucleon states,
\begin{align}
 f^N_q m_N &= \bra{N}m_q \bar q q \ket{N} , \\
 -\frac{8}{9}f^N_g m_N &= \bra{N}\frac{\alpha_s}{\pi} G^a_{\mu\nu}G^{a \mu\nu} \ket{N}.
\end{align}
Their approximate values are given as \cite{Hisano:2015rsa}
\begin{equation}
 f_u^p = 0.019, \quad f_d^p = 0.027, \quad f_u^n = 0.013, \quad f_d^n = 0.040, \quad  f_s^p=f_s^n = 0.009,
\end{equation}
\begin{equation}
 f_g^N  = 1-f_u^N-f_d^N-f_s^N + \mathcal{O} (\alpha_s).
\end{equation}

\begin{figure}[tbp]
\centering
\includegraphics[width=0.3\textwidth]{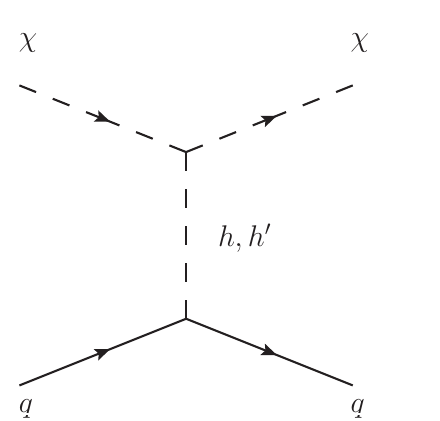} \hspace{5mm}
\includegraphics[width=0.3\textwidth]{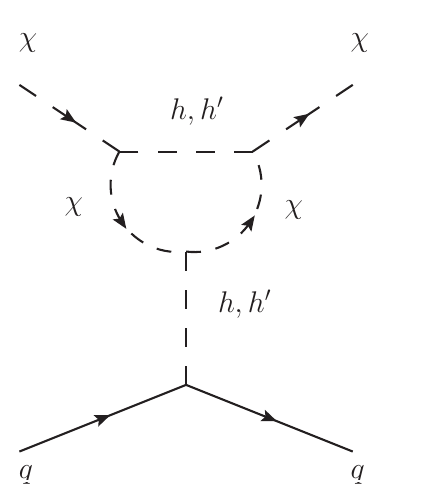}
\caption{
\textbf{(Left:)} A tree-level scattering process between $\chi$ and quarks.
\textbf{(Right:)} An example of loop diagrams that include the tree-level one as its sub diagram.
It vanishes due to the same mechanism as the tree-level one.
}
\label{fig:cancel-diagram}
\end{figure}

\begin{figure}[tbp]
\centering
\includegraphics[width=0.3\textwidth]{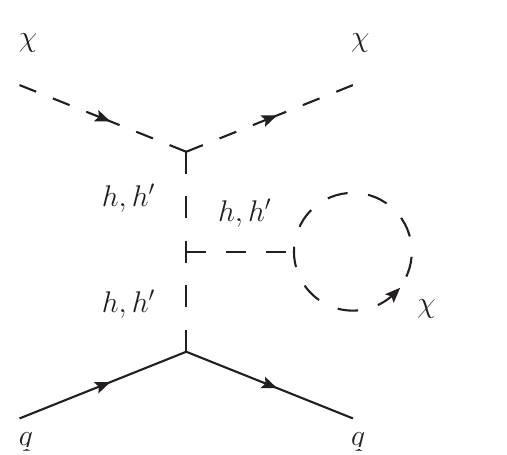} \hspace{5mm}
\includegraphics[width=0.27\textwidth]{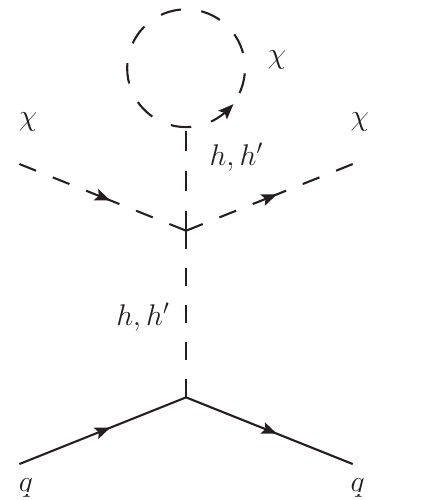} \hspace{5mm}
\includegraphics[width=0.27\textwidth]{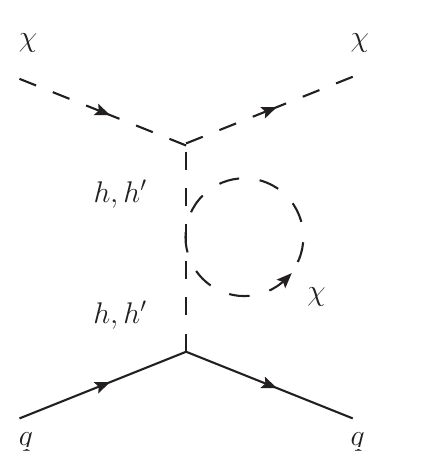} 
\caption{
An example of non-trivial cancellation.
Although these three diagrams depend on $m_\chi$ explicitly, they are canceled and vanish with non-zero $m_\chi$.
}
\label{fig:cancel-tadpole-diagram}
\end{figure}

\begin{figure}[tbp]
\centering
\includegraphics[width=1\textwidth]{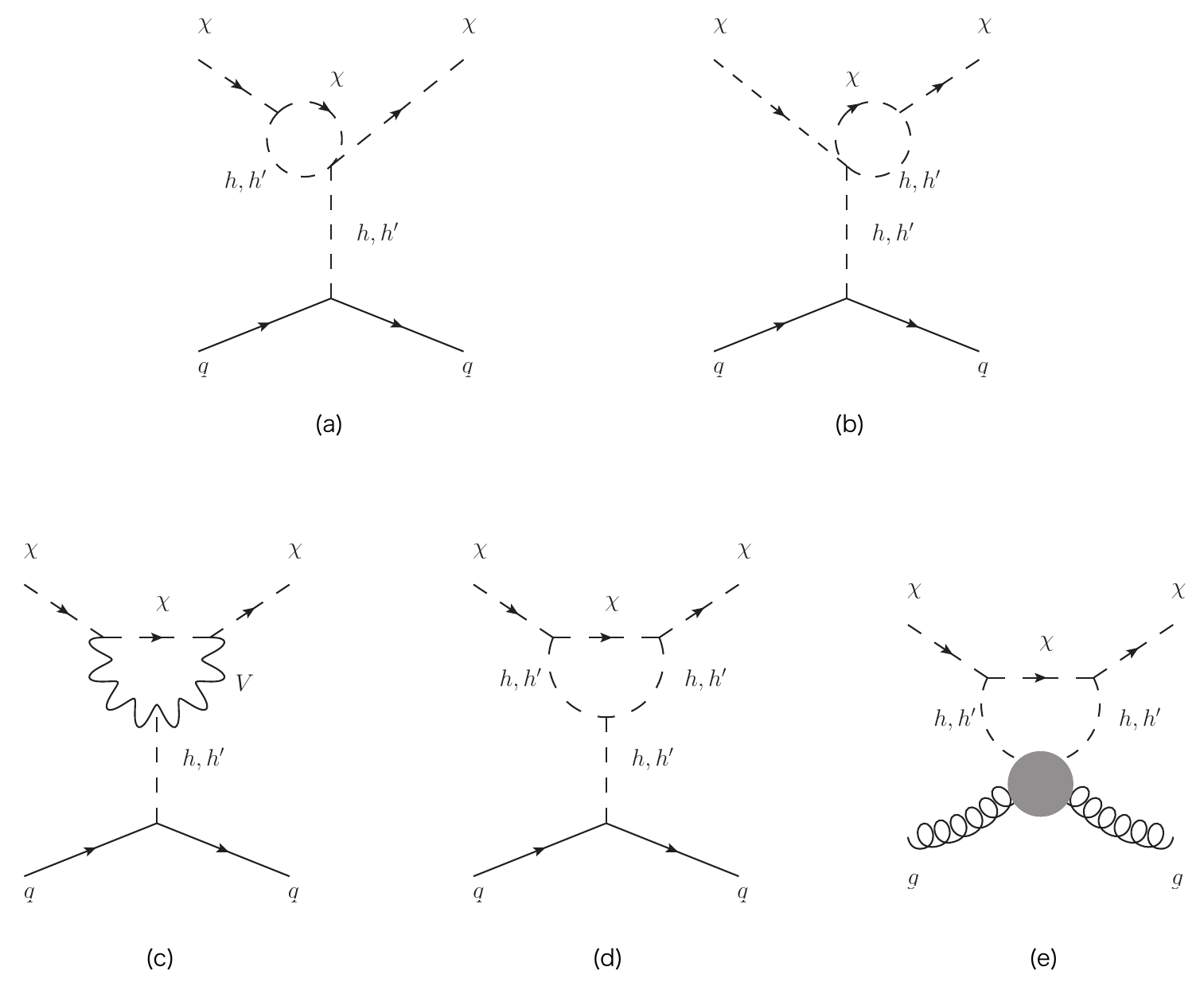} 
\caption{
Feynman diagrams relevant for the scattering with nucleons.
The gray blob in the last diagram (e) indicates the QCD NLO contribution, which is dominated by the top-quark loop~\cite{Ishiwata:2018sdi}.
}
\label{fig:feynman-diagram}
\end{figure}

Now our task is to calculate the scattering amplitude with quarks and gluon at the one-loop (and QCD NLO) level taking the momentum transfer to be zero ($t \to 0$), and then to read off the Wilson coefficients from them.
Note that some diagrams include the tree-level process as their sub diagrams, and thus vanish due to the cancellation of $h$ and $h'$ exchanges (see Fig.~\ref{fig:cancel-diagram}).
We evaluate the remaining diagrams in terms of the following technique.
In the limit $m_\chi\to 0$ where the pNG boson becomes an exact Nambu-Goldstone boson,
all diagrams cancel and the total scattering amplitude vanishes as $t \to 0$ even at the loop level~\cite{Azevedo:2018exj, Glaus:2020ihj},
\begin{equation}
i \mathcal{M}_\mathrm{tot}(m_\chi) \equiv \sum_\mathrm{diagrams} i \mathcal{M}(m_\chi) \to 0 \quad (m_\chi\to 0).
\end{equation}
On the other hand, the cancellation among the diagrams is not exact when $m_\chi \neq 0$.
Utilizing this property, we can rewrite the amplitudes as 
\begin{equation}
i \mathcal{M}_\mathrm{tot}(m_\chi) = i \mathcal{M}_\mathrm{tot}(m_\chi) - i \mathcal{M}_\mathrm{tot}(0) = \sum_\mathrm{diagrams} \left[i \mathcal{M}(m_\chi) - i \mathcal{M}(0) \right] ,
\end{equation}
from which it follows that one does not need to calculate diagrams that are independent of $m_\chi$
since they are canceled within the bracket.
In addition, tadpole diagrams that depend on $m_\chi$ are cancelled by a self-energy diagram, see Fig.~\ref{fig:cancel-tadpole-diagram}.
Thus all diagrams we need to evaluate are five diagrams shown in Fig.~\ref{fig:feynman-diagram}:
\begin{equation}
 i \mathcal{M}_\mathrm{tot}(m_\chi) = \sum_\text{diagrams in Fig.~\ref{fig:feynman-diagram}} \left[i \mathcal{M}(m_\chi) - i \mathcal{M}(0) \right] .
\end{equation}

\begin{figure}[tbp]
\centering
\includegraphics[width=0.45\textwidth]{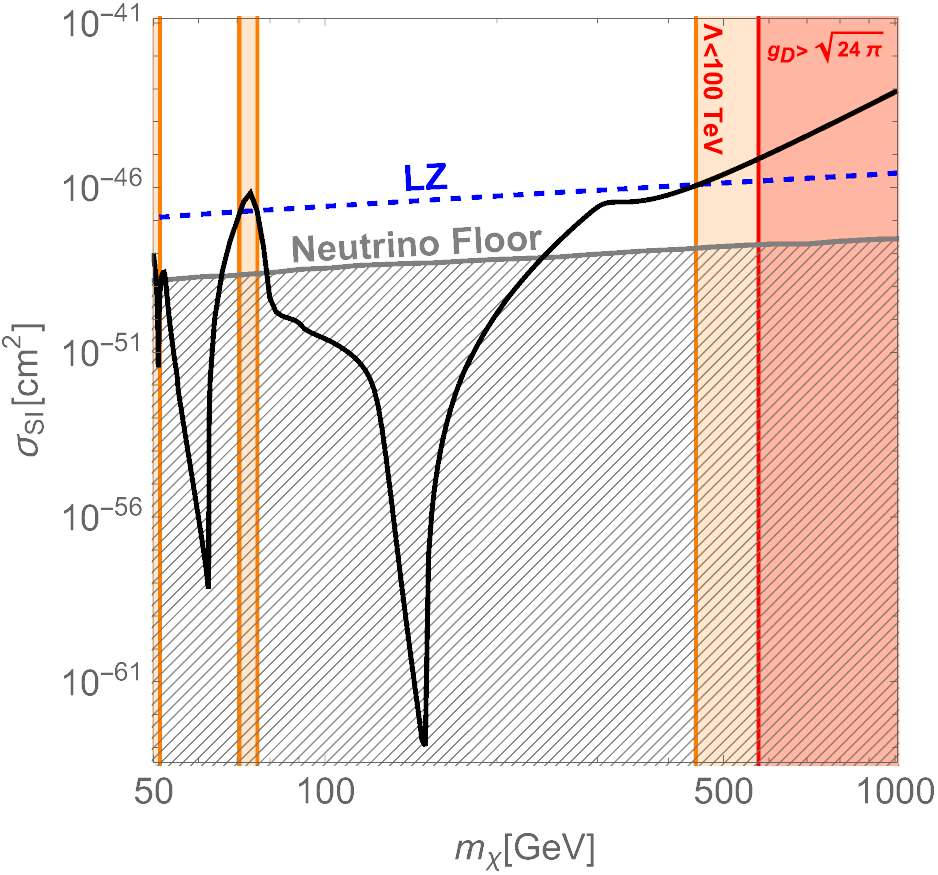} \hspace{2mm}
\includegraphics[width=0.45\textwidth]{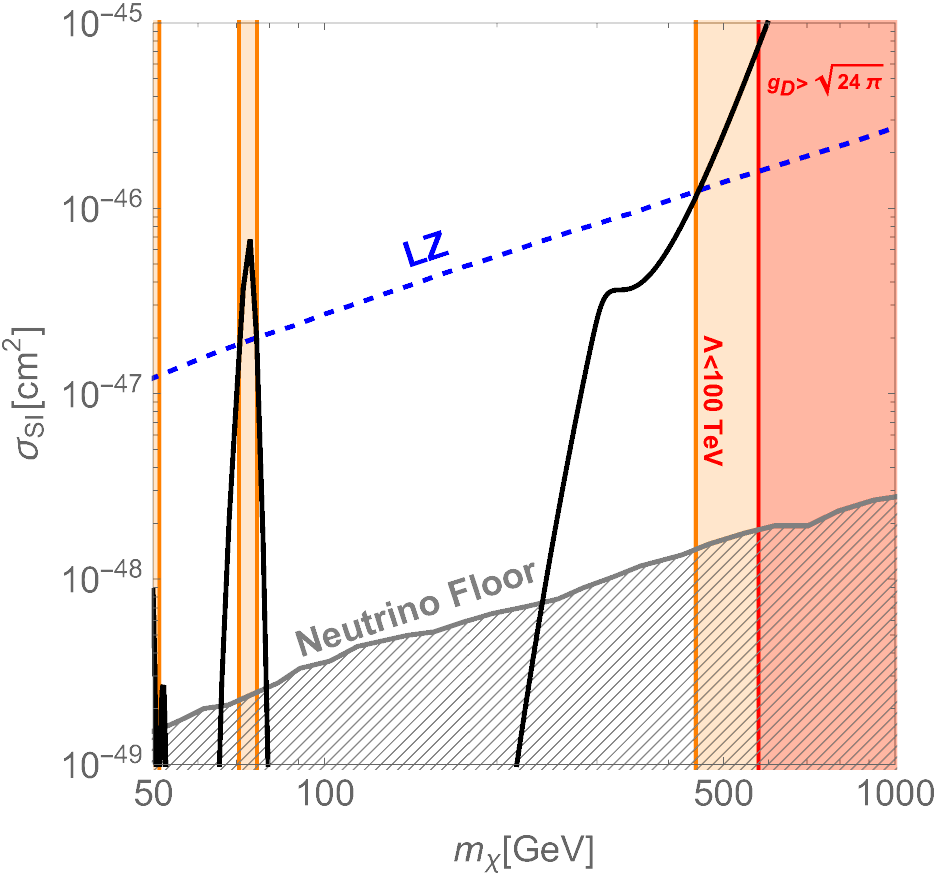}
\caption{
Plots of the DM mass $m_\chi$ vs. the SI scattering cross section $\sigma_{SI}$ between the pNG DM $\chi$ and nucleons.
On the black solid line, the DM relic abundance is explained by the thermal freeze-out mechanism.
The red shaded region is excluded by the perturbative unitarity of $g_D$.
In the orange shaded regions, the running coupling constant $g_D(\Lambda)$ violates the perturbative unitarity below $\Lambda = 100$ TeV.
The current upper bound by the LZ experiment is shown by the blue dashed line.
The gray shaded region indicates the cosmic neutrino background.
The right panel is zoomed-up version of the left one.
}
\label{fig:SI-cross-section}
\end{figure}

\begin{figure}[tbp]
\centering
\includegraphics[width=0.45\textwidth]{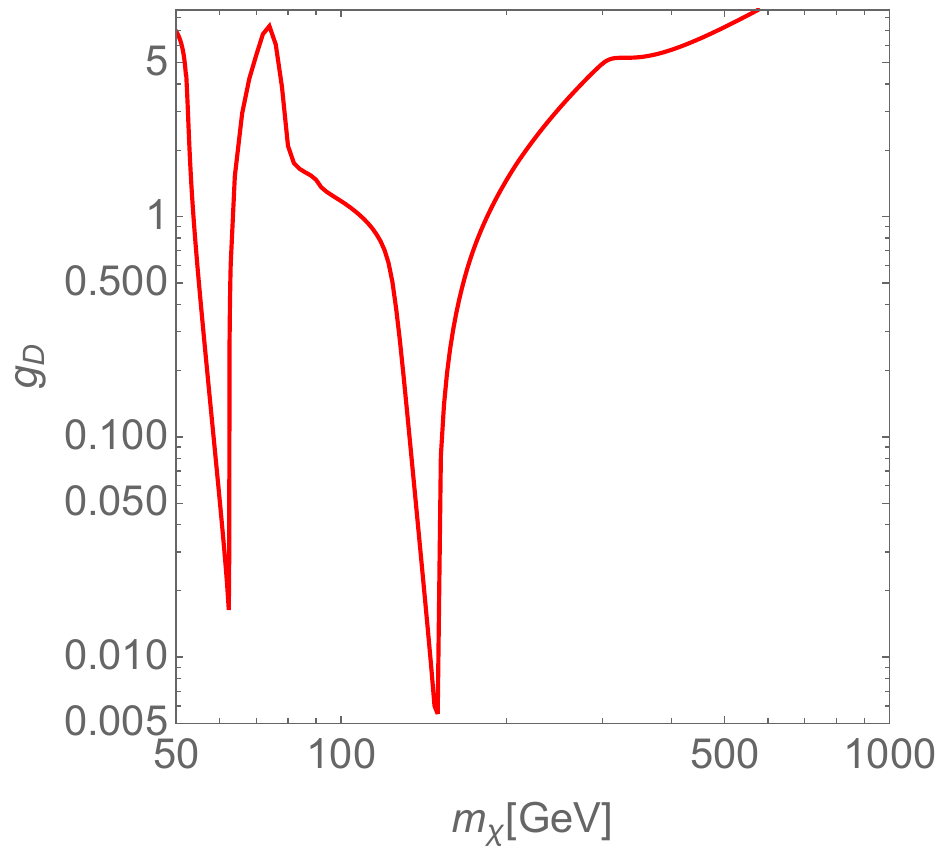} \hspace{2mm}
\caption{
The value of $g_D$ for a given parameter set. 
Here we take $\sin\theta = 0.1, m_V = 3 m_\chi$, and $v_s$ is given in 
Fig.~\ref{fig:omegah2}.
}
\label{fig:gD}
\end{figure}

The explicit expressions of the amplitudes and the Wilson coefficients are shown in Appendix~\ref{sec:appendix2}.
We only show the numerical results of the SI cross section $\sigma_{SI}$ by Fig.~\ref{fig:SI-cross-section}.
The parameters are taken as $m_{h'}=300\, \mathrm{GeV}$, $m_{V}=3 m_\chi$, $\sin \theta=0.1$,
and $v_s$ is determined such that the DM relic abundance is explained by the thermal freeze-out mechanism as studied in Sec.~\ref{sec:relic}.
The red shaded region is excluded by the violation of the perturbative unitarity of the $U(1)_D$ gauge coupling constant, $g_D > \sqrt{24 \pi}$, see Eq.~\eqref{eq:PU-gauge}.
In the orange shaded regions, the running coupling constant $g_D(\Lambda)$ calculated by the one-loop beta function exceeds $\sqrt{24 \pi}$ below $\Lambda = 100$ TeV.
The current bound by the LZ experiment~\cite{LUX2022} is shown by the blue dashed line.
The gray hatched region indicates that the DM signal is hidden by the neutrino background.
We find that $\sigma_\text{SI}$ is smaller than the neutrino floor in most of the region. For the larger $m_\chi$ regime, $\sigma_\text{SI}$ is larger than the neutrino floor and some region is already excluded. 
This is because, in that region, the gauge coupling constant becomes large as seen in Fig.~\ref{fig:gD}. 
The diagram containing the gauge boson (Fig.~\ref{fig:feynman-diagram}-(c)) is proportional to $g_D^4$,
and hence gives a significantly large contribution.
We conclude that, in almost all parameter space, the SI cross section $\sigma_{SI}$ is far below the current experimental bound as long as we keep the value of $g_D$ small.

\section{two-component scenario}\label{sec:two-component}

Here we discuss the two-component scenario in this model where $m_V < 2 m_\chi$.
In this scenario, the DM energy density is the sum of the energy densities of $\chi$ and $V$, namely $\Omega h^2 = (\Omega h^2)_\chi + (\Omega h^2)_V$. 
We determine the value of $v_s$ to obtain the right amount of the DM energy density.

\begin{figure}
\centering
\includegraphics[width=0.48\hsize]{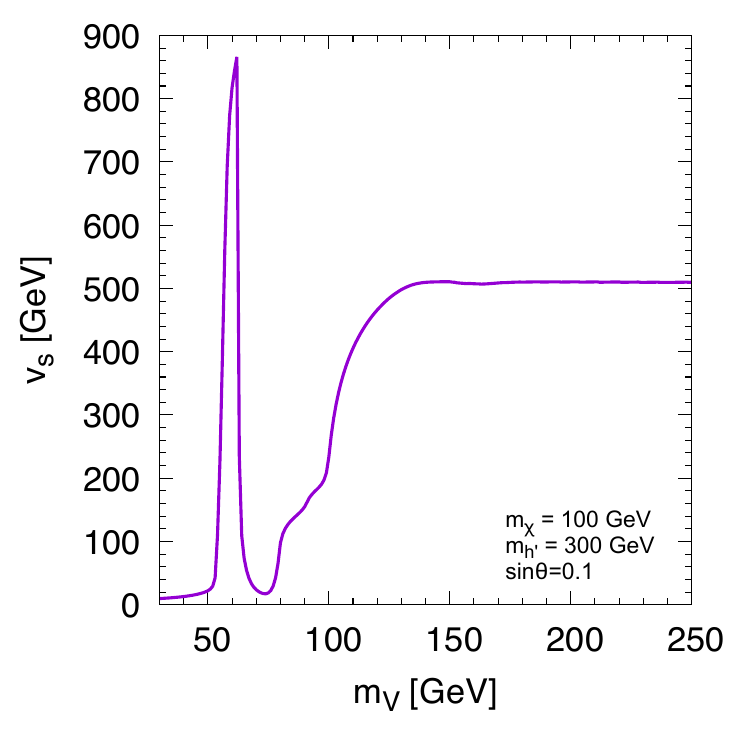}
\includegraphics[width=0.48\hsize]{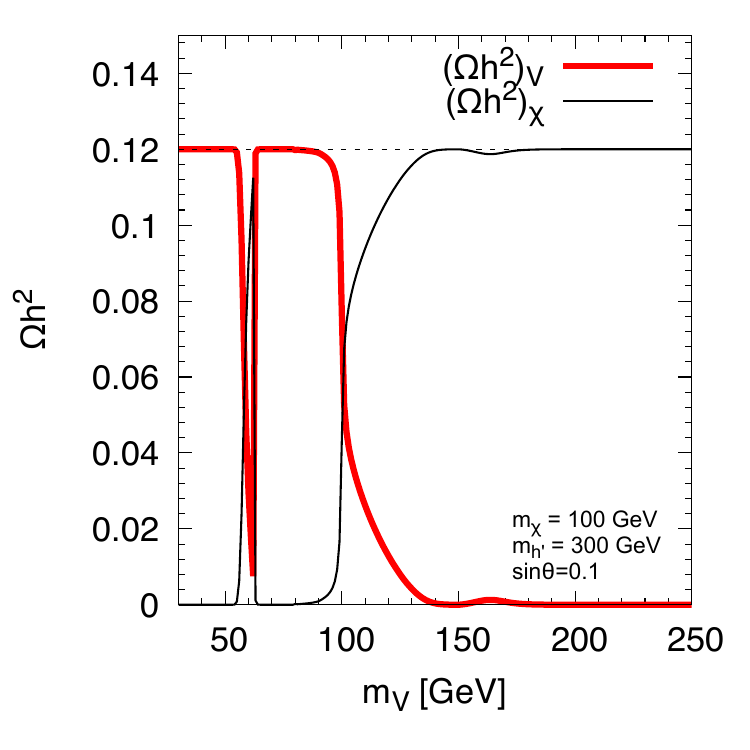}
\includegraphics[width=0.48\hsize]{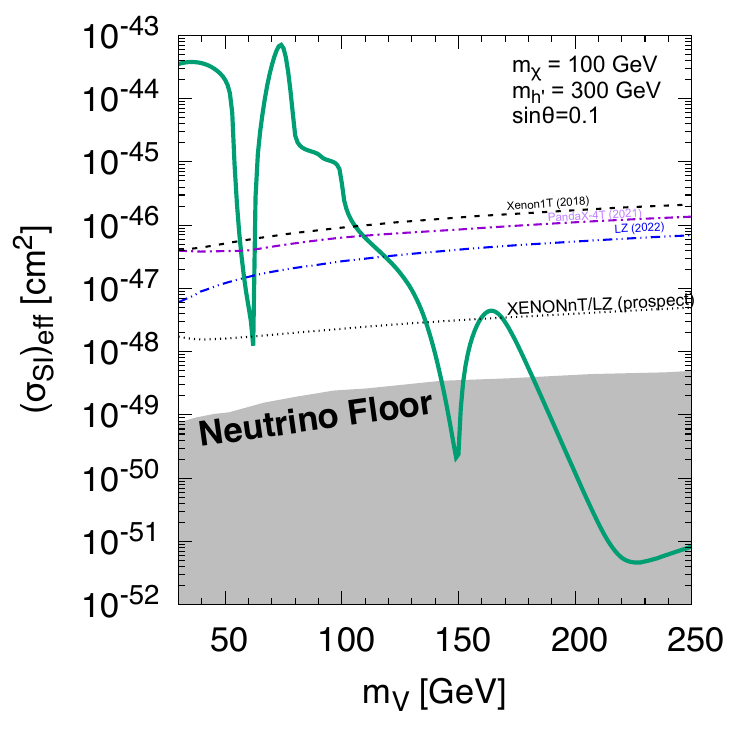}
\includegraphics[width=0.48\hsize]{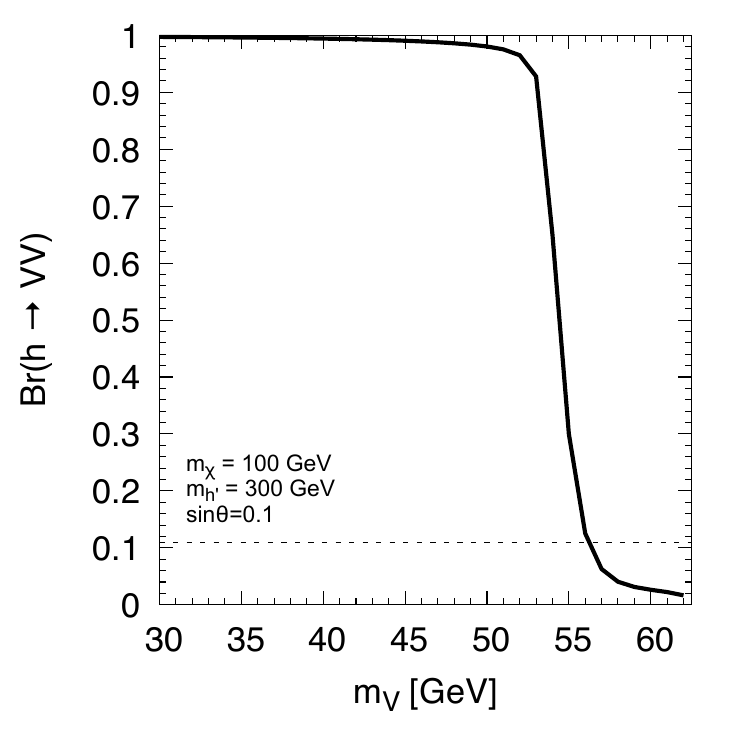}
\caption{
Results in the two-component DM scenario. Here we choose $m_\chi= 100$~GeV, $m_{h'}=300$~GeV, and $\sin\theta = 0.1$ as a reference value.
\textbf{(top-left):} the value of $v_s$ that reproduces the measured value of the DM energy density.
\textbf{(top-right):} the relic abundance of $V$ and $\chi$.
\textbf{(bottom-left):} The green-solid-curve shows the model prediction of $\sigma_\text{SI}$ scaled by the fractions of the relic abundances of $V$ and $\chi$.
\textbf{(bottom-right):} The branching ratio of $h \to VV$. The horizontal dashed line shows the current upper bound on the branching ratio.
}
\label{fig:two-component}
\end{figure}
The top-left panel in Fig.~\ref{fig:two-component} shows the value of $v_s$ that reproduces $\Omega h^2 = 0.12$. Here we choose $m_\chi= 100$~GeV, $m_{h'}=300$~GeV, and $\sin\theta = 0.1$ as a reference parameter set and vary the value of $m_V$.
For $m_V > 2 m_\chi$, $V$ can decay into $\chi$ and thus $\Omega h^2$ is determined by the abundance of $\chi$. Hence $\Omega h^2$ is independent of $m_V$.
For $m_V < 2 m_\chi$, $V$ is also a stable particle. Thus the relic abundance of $V$ also contributes to $\Omega h^2$, and $v_s$ depends on the value of $m_V$.
We use this value of $v_s$ for all the panels in Fig.~\ref{fig:two-component}.

The top-right panel in Fig.~\ref{fig:two-component} shows the value of the relic abundance of $V$ and $\chi$. We denote them $(\Omega h^2)_V$ and $(\Omega h^2)_\chi$, respectively.
We find that $(\Omega h^2)_V > (\Omega h^2)_\chi$ for $m_\chi < m_V$.
The exception is found around $m_V = m_h/2 \simeq 62.5$~GeV. This is due to the Higgs resonant enhancement in the annihilation of $V$. In that region of the parameter space, a $VV$ pair efficiently annihilate into the SM particles by exchanging the SM Higgs boson in the $s$-channel, and thus $\chi$ is the dominant component of DM. 

As we discussed in Sec.~\ref{sec:scattering-amplitude}, the $V$-nucleon scattering process is not suppressed by the small momentum transfer.
Hence the two-component scenario is testable through the direct detection experiments.
The experiments give the upper bound on the spin-independent cross section $\sigma_\text{SI}$ under the assumption that the DM is single component.
To compare the two-component scenario with the experiments, we define the effective spin-independent cross section by scaling according to the fraction of the $V$ and $\chi$ components,
\begin{align}
 (\sigma_{\text{SI}})_\text{eff}  = \sigma_{\text{SI}}^{V} \frac{(\Omega h^2)_V}{(\Omega h^2)_V+(\Omega h^2)_\chi}
 +
 \sigma_{\text{SI}}^{\chi} \frac{(\Omega h^2)_\chi}{(\Omega h^2)_V+(\Omega h^2)_\chi},
\end{align}
where $\sigma_{\text{SI}}^{V} $ and $\sigma_{\text{SI}}^{\chi}$ are the cross section for $VN \to VN$ and $\chi N \to \chi N$, respectively. The latter is suppressed by the small momentum transfer but is generated at the loop level as discussed in Sec.~\ref{sec:loop}.
The bottom-left panel in Fig.~\ref{fig:two-component} shows $(\sigma_{\text{SI}})_\text{eff}$.
Comparing to the recent results from the direct detection experiments~\cite{1805.12562, PandaX-4T:2021bab, LUX2022}, 
we find that the region for $m_V \lesssim 120$~GeV except for the Higgs resonant region is excluded.
This is a natural consequence because the $V$-nucleon scattering is not suppressed by the small momentum transfer. From this result and the top-right panel, 
we find that $V$ cannot be the dominant component of DM in the two-component scenario in this model.

We also show Br$(h \to VV)$ in the bottom-right panel in Fig.~\ref{fig:two-component}. 
The partial width of $h \to VV$ is given by
\begin{align}
\Gamma(h \to VV) = \frac{1}{32\pi m_h} g_{hVV}^2 \left( 2 + \frac{(m_h^2 -2 m_V^2)^2}{4 m_V^4}\right) \sqrt{1 - \frac{4 m_V^2}{m_h^2}}
.
\end{align}
This can be measured at the collider experiments as the Higgs invisible decay.
We find that $m_V \lesssim 56$~GeV is excluded.

\section{Summary}\label{sec:summary}
We have proposed a new pNG DM model that can explain the null results of the direct detection experiments. 
There is an $SU(2)_g$ global symmetry and a $U(1)_X$ gauge symmetry in the dark sector. 
We assume that the VEV of a complex scalar field $\phi$ breaks these symmetries into $U(1)_D$ global symmetry spontaneously.
We also assume that the $SU(2)_g$ global symmetry is explicitly and softly broken into $U(1)_g$. 
Under the setup, we obtain a complex pNG scalar ($\chi$) that is stable thanks to the $U(1)_D$ global symmetry. 
This $\chi$ is a DM candidate in this model.
We also impose that the dark sector has a discrete symmetry defined in
Eqs.~\eqref{eq:dark-charge-conjugation-1} and \eqref{eq:dark-charge-conjugation-2}, 
which can be identified as the charge conjugation in the dark sector. This discrete symmetry forbids the gauge kinetic mixing between the $U(1)_X$ in the dark sector and the $U(1)$$_Y$ in the visible sector. 
The $U(1)_X$ gauge boson can be stable thanks to this discrete symmetry if its mass is less than twice the pNG DM mass.

The advantage of our model is that the soft symmetry breaking terms other than $\phi^\dagger \tau^3 \phi$ are forbidden by the $U(1)_X$ gauge symmetry. 
This restriction on the soft symmetry breaking term guarantees the suppression in the scattering amplitudes by the momentum transfer.

We have calculated the relic abundance and determined the DM coupling to the Higgs boson that reproduces the measured value of the DM energy density. Some region of the parameter space requires relatively large coupling, 
and thus we also have checked the perturbative unitarity bound.
We have found that most of the parameter space is consistent with the PU bound. 
It is also found that the $U(1)_X$ gauge coupling can be large and the heavy pNG DM region is excluded by the PU bound if we assume the mass of the gauge boson is larger than twice the mass of the pNG DM. 
In addition, we have considered the cosmic string in our model, which is a softly-broken version of the semilocal string and is topologically stable due to the non-trivial topology of the vacuum.
We have checked that the DM contribution from the cosmic string loops is insignificant. Moreover, we have discussed the loop effect on the spin-independent cross section. We have found that the effect is small unless the $U(1)_X$ gauge coupling is large.

The dark sector contains the gauge boson as well as the pNG DM. If the mass of the gauge boson is smaller than twice the pNG DM mass, then the gauge boson is also stable due to the discrete symmetry (charge conjugation) in the dark sector. Therefore the model contains the two-component DM scenario depending on the mass spectra. 
We briefly discussed the two-component DM scenario in Sec.~\ref{sec:two-component}. 
We showed that the results in the direct detection experiments give strong bound on the scenario because $V$ is not pNG DM and thus the $V$-nucleon scattering is not suppressed by the small momentum transfer.

The model contains the $U(1)_X$ gauge symmetry, which is not asymptotic free and would hit the Landau pole at high energy. We need a small gauge coupling or some UV completion. One possible extension of the current model is to embed the $U(1)_X$ gauge symmetry into an $SU(2)$ gauge symmetry that is broken into the $U(1)_X$ gauge symmetry spontaneously. 
Discussion of the possible UV picture will be discussed in the future.

\section*{Acknowledgments}
The authors thank Christian Gross for a valuable comment.
Y.~H. would like to thank Yoshihiko Abe for useful discussion.
This work is supported in part by JSPS KAKENHI Grant Numbers 19H04615 and 21K03549 (TA)
and JP21J01117 (YH).
The work is also supported by JSPS Core-to-Core Program (grant number:JPJSCCA20200002).


\appendix
\section{Cosmic string solution}
\label{sec:appendix}
We here discuss the solution of the cosmic string in our model.
For simplicity, we concentrate on the dark sector consisting of $\phi$ and $V_\mu$ and ignore their portal coupling with the SM sector, $\lambda_{H\phi}=0$.
This would not change the following argument much.
The effect of such a portal coupling is studied in Ref.~\cite{Hyde:2013fia}.

Firstly, we rewrite the Lagrangian of the dark sector in Eq.~\eqref{eq:lagrangian} as
\begin{align}
 \mathcal{L}_\text{dark} = |D_\mu \phi|^2 - \frac{1}{4} V^{\mu\nu} V_{\mu\nu} - \lambda_\phi\left(|\phi|^2 - \frac{v_s^2}{2}\right)^2 - m_\chi^2 \phi_1^2
\end{align}
with $\phi=(\phi_1, \phi_2)^t$.
To discuss the string solutions, it is convenient to rescale the fields as
\begin{equation}
 V_\mu \to \frac{2}{g_D} V_\mu , \quad \phi \to \frac{2}{g_D} \phi,
\end{equation}
leading to 
\begin{align}
 \mathcal{L}_\text{dark} = \frac{4}{g_D^2} \left[|\hat D_\mu \phi|^2 - \frac{1}{4} V^{\mu\nu} V_{\mu\nu} - \frac{\beta}{2}\left(|\phi|^2 - \frac{g_D^2 v_s^2}{8}\right)^2 - m_\chi^2 \phi_1^2 \right]
\end{align}
with $\hat D_\mu = \partial_\mu + i V_\mu$.
We have introduced a parameter $ \beta \equiv 8\lambda_\phi/g_D^2$, which is equal to the squared mass ratio of the gauge and scalar bosons, $\beta = m_{h'}^2/m_V^2$.
In the following, we take the dimensionless unit such that $g_D v_s/2 =1$,
which is equivalent to introducing the dimensionless variables (denoted by tilde)
\begin{equation}
 V_\mu = \frac{g_D v_s}{2} \tilde V_\mu, \quad x^\mu = \tilde x^\mu \frac{2}{g_D v_s}, \quad \phi = \frac{g_D v_s}{2} \tilde \phi, \quad m_{\chi} = \frac{g_D v_s}{2}\tilde m_{\chi},
\end{equation}
for which the Lagrangian is given by 
\begin{align}
  \mathcal{L}_\text{dark} = g_D^2 v_s^4\left[|\tilde D_\mu \tilde \phi|^2 - \frac{1}{4} \tilde{V}^{\mu\nu} \tilde{V}_{\mu\nu} - \frac{\beta}{2}\left(|\tilde \phi|^2 - \frac{1}{2}\right)^2 - \tilde m_\chi^2 \tilde \phi_1^2 \right]
\end{align}
with $\tilde D_\mu = \tilde \partial_\mu + i \tilde V_\mu$.
In the following, we drop the tildes for simple notation.

We consider the axially symmetric ansatz for the string solution,
\begin{align}
 \phi = \frac{1}{\sqrt{2}}
\begin{pmatrix}
 h(r) \\ f(r) e^{ i\theta}
\end{pmatrix}, \quad
V_\theta = - ( 1- w(r))\label{eq:ansatz}
\end{align}
and $V_r = V_z = V_t = 0$.
We have introduced the cylinder coordinate $x+ iy = r e^{i\theta}$.
The profile functions, $f(r), h(r)$ and $w(r)$, should satisfy the boundary conditions
\begin{align}
& f(0)= 0,\quad \quad h'(0) = 0, \quad w(0)=1, \\
& f(\infty) = 1, \quad h(\infty)=0, \quad w(\infty)=0 \, .
\end{align}
Since this configuration behaves asymptotically at $r\to  \infty$ as
\begin{equation}
 \phi \sim \frac{1}{\sqrt 2} e^{i \theta}
\begin{pmatrix}
0 \\ 1
\end{pmatrix}, \quad 
V_\theta \sim - 1,\label{eq:asymp}
\end{equation} 
it has the unit winding number in the vacuum manifold $U(1)_g\times U(1)_X/U(1)_D$,
and hence describes a topologically stable solution. 
The profile functions are determined by solving the classical equations of motion (EOMs).

Before solving the EOMs,
let us consider the so-called semilocal model~\cite{Vachaspati:1991dz,Achucarro:1999it}, which is equivalent to our model with the soft symmetry breaking term chopped off, $m_\chi \to 0$.
There, the symmetry breaking pattern is $SU(2)_g\times U(1)_X \to U(1)_D$,
and the existence of the stable string solution (called semilocal string) depends on the parameter $\beta$~\cite{Hindmarsh:1991jq,Achucarro:1992hs}.
This is because of the trivial first homotopy group, $\pi_1(SU(2)_g\times U(1)_X /U(1)_D) \simeq 0$.
For $\beta \leq 1$, the string is classically stable (i.e., local minimum of the energy) and is nothing but the embedding of the  Abrikosov-Nielsen-Olesen (ANO) string~\cite{Abrikosov:1956sx,Nielsen:1973cs} into $\phi_2$ and $V_\mu$ with  $\phi_1$ vanishing everywhere.
On the other hand, for $\beta>1$, it is unstable because a condensation of $\phi_1$ happens inside the core of the string and the condensed region expands to outward infinitely, resulting in no static stable configuration.
Thus the critical value of $\beta$ for the stability is equal to the value that is the boundary for whether the condensation of $\phi_1$ occurs or not (we denote it by $\beta_b$ and $\beta_b=1$ in this case).

Let us move on to our model, $m_\chi \neq 0$.
In this case, $m_\chi$ gives a positive constant mass for $\phi_1$ and the condensation of $\phi_1$ must be suppressed.
Thus the boundary value of $\beta$ for the condensation, $\beta_b$, must be shifted to be larger than unity $\beta_b>1$ depending on $m_\chi$.
Furthermore, even when the condensation occurs for $\beta> \beta_b$, 
it does not expand infinitely but stops at finite size $\sim m_\chi^{-1}$ because $\phi_1$ must be zero at large distances from the string core due to the positive mass, as Eq.~\eqref{eq:asymp}.
This means that the condensation of $\phi_1$ does not destabilize the string solution,
which is consistent with the topological argument $\pi_1(U(1)_g\times U(1)_X /U(1)_D) \simeq \mathbb{Z}$.

\begin{figure}[tbp]
\centering
\includegraphics[width=0.45\textwidth]{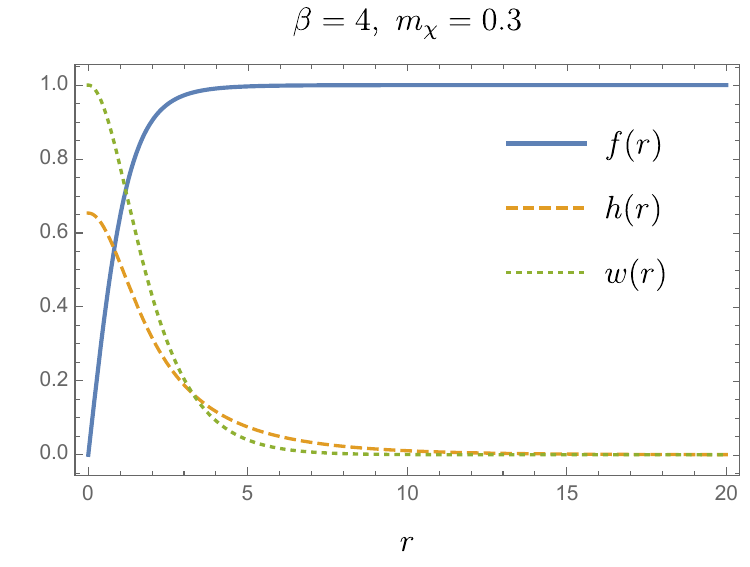} \hspace{1em}
\includegraphics[width=0.45\textwidth]{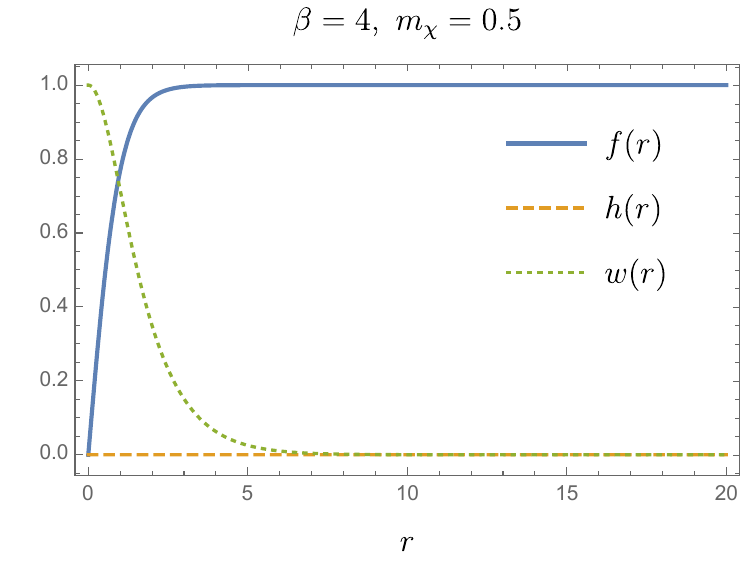} \\[2ex]
\includegraphics[width=0.45\textwidth]{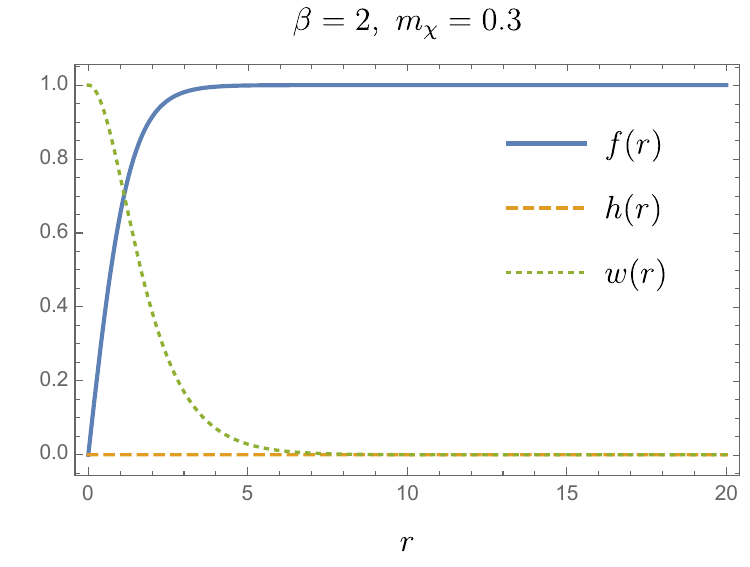}
\caption{
Plots for the string solutions.
All dimensionful quantities are normalized by $g_D v_s/2$.
In the left-top panel, the condensation of $\phi_1= h(r)/\sqrt{2}$ takes place in the string core.
As $\beta$ decreases or $m_\chi$ increases, the condensation is suppressed, as in the right-top and bottom panel.
}
\label{fig:string}
\end{figure}

\begin{figure}[tbp]
\centering
\includegraphics[width=0.5\textwidth]{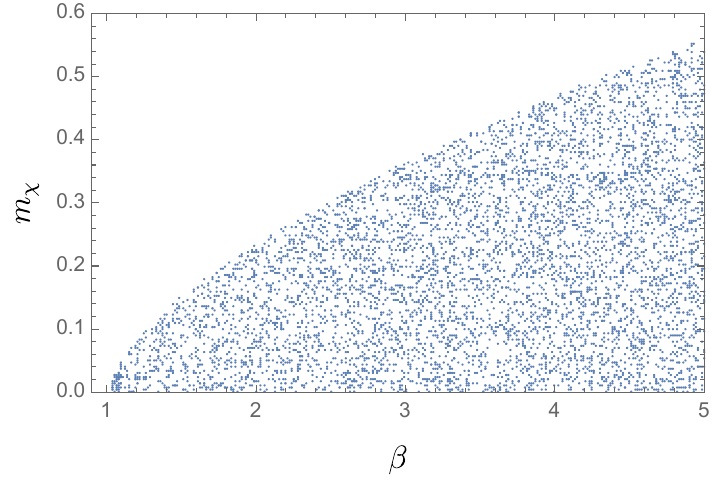} 
\caption{
Scatter plot for the parameter region $(\beta , m_\chi)$ in which the condensation of $\phi_1$ occurs in the string core.
$m_\chi$ is normalized by $g_D v_s/2$.
}
\label{fig:condense}
\end{figure}

Substituting the ansatz \eqref{eq:ansatz}, we obtain the EOMs as
\begin{align}
 f'' + \frac{f'}{r} - \frac{w^2}{r^2} f - \frac{\beta}{2} f\left(f^2+h^2-1 \right) & = 0, \\
 h'' + \frac{h'}{r} - \frac{(w-1)^2}{r^2} h - m_\chi h  - \frac{\beta}{2} h\left(f^2+h^2-1 \right) & = 0, \\ 
 w'' - \frac{w'}{r} -  (f^2+h^2) w + h^2 & = 0 .
\end{align}
We solve the EOMs numerically by the relaxation method and show the string solutions in Fig.~\ref{fig:string}.
Note that all dimensionful quantities are normalized by $g_D v_s/2$.
In the left-top panel, the condensation of $\phi_1= h(r)/\sqrt{2}$ takes place in the string core.
As $\beta$ decreases or $m_\chi$ increases, the condensation is suppressed, as in the right-top and bottom panels.
In the latter case, the solution is the same as the standard ANO vortex string.
Fig.~\ref{fig:condense} shows a parameter region in which $\phi_1$ condenses inside the string core.
We can see that a quite small value of $m_\chi$ suffices to suppress the condensation.
Around $m_\chi\simeq 0$, the condensation can occur only for $\beta>1$, which is consistent with the semilocal case.
Note that all solutions even with the condensation are topologically stable thanks to the non-trivial winding number (Eq.~\eqref{eq:asymp}).

The above result is similar to the one in Ref.~\cite{Eto:2016mqc} that admits semilocal strings with an explicit $SU(2)$-breaking term.
There, the $SU(2)$-breaking term is given as a mass-dimension-four operator, $\left(\phi^\dagger T^3 \phi\right)^2$, and both of $\phi_1$ and $\phi_2$ develop their VEVs at the vacuum.
As a result, the solution is topologically stable and can be the ``vortex molecule'' consisting of two half-quantized strings depending on the parameters.
On the other hand, in our model, $\phi_1$ vanishes outside the string core due to the soft breaking term $\phi^\dagger T^3 \phi$.
Thus the string solution is qualitatively different from that in Ref.~\cite{Eto:2016mqc}.

One may consider that the nonzero $\phi_1$ in the string solution breaks the $U(1)_D$ symmetry, making it the superconducting string~\cite{Witten:1984eb}.
However, this is not the case.
In the soliton background, the generator of the $U(1)_D$ symmetry should be defined as $-T^a n^a + 1/2$ with $n^a = (\phi^\dagger \tau ^a \phi)/|\phi|^2$.
Thus the solution given as Eq.~\eqref{eq:ansatz} is always invariant under this symmetry.

\section{Scattering amplitudes of $\chi$ and nucleons}
\label{sec:appendix2}
We here provide the explicit expressions of the scattering amplitudes shown in Fig.~\ref{fig:feynman-diagram}.
In the following, we denote $h_1 \equiv h$, $h_2 \equiv h'$.
It is straight forward to calculate the diagrams (a)-(d):
\begin{align}
 i \mathcal{M}_a = i \mathcal{M}_b = -\bar u u \sum_{j,k=1,2} g_{ff h_j} \frac{1}{m_{h_j}^2} g_{\chi\chi h_j h_k} g_{\chi \chi h_k} \frac{i}{(4\pi)^2} B_0(m_\chi^2, m_{h_k}^2, m_\chi^2) \, ,
\end{align}

\begin{align}
 i \mathcal{M}_c &= \bar u u \sum_{j=1,2} \frac{g_{ff h_j} g_{VVh_j}}{m_{h_j}^2} \frac{g_D^2}{4} \frac{i}{(4\pi)^2} 
\left[ (-m_V^2 + 4 m_\chi^2) \frac{ \partial}{\partial m_V^2}B_0(m_\chi^2, m_V^2, m_\chi^2) \right. \nonumber \\
& \left. \phantom{ (-m_V^2 + 4 m_\chi^2) \frac{ \partial}{\partial m_V^2}B_0(m_\chi^2, m_V^2, m_\chi^2)} 
-B_0(m_\chi^2, m_V^2, m_\chi^2) + B_0(0, m_V^2, m_V^2)\right]\, ,
\end{align}

\begin{align}
 i \mathcal{M}_d = -\bar u u \sum_{j,k,l=1,2} \frac{g_{ff h_j} g_{\chi\chi h_j} g_{\chi \chi h_k} g_{h_j h_k h_l}}{m_{h_j}^2} \frac{i}{(4\pi)^2} C_0(m_\chi^2,m_\chi^2, 0,m_k^2,m_\chi^2,m_l^2)\, ,
\end{align}
where $u$ is the spinor wavefunction and 
\begin{equation}
 \frac{i}{(4\pi)^2} B_0(p^2, m_1^2, m_2^2) \equiv \int \frac{\dd ^d l}{(2\pi)^d} \frac{1}{(l^2 - m_1^2)(l+p)^2 - m_2^2} ,
\end{equation}
\begin{equation}
 \frac{i}{(4\pi)^2} C_0(p_1^2,p_2^2,(p_1+p_2)^2, m_1^2, m_2^2,m_3^2) \equiv \int \frac{\dd ^d l}{(2\pi)^d} \frac{1}{(l^2 - m_1^2)[(l+p_1)^2 - m_2^2][(l+p_1+p_2)^2 - m_3^2]},
\end{equation}
\begin{align}
g_{\chi\chi hh} =& 2 \lambda_\phi s_\theta^2 + \lambda_{H\Phi} c_\theta^2,\\
g_{\chi\chi hh'} =& -2 \lambda_\phi s_\theta c_\theta + \lambda_{H\Phi} s_\theta c_\theta,\\
g_{\chi\chi h'h'} =& 2 \lambda_\phi c_\theta^2 + \lambda_{H\Phi} s_\theta^2,\\
g_{hhh}=& 6 v \lambda_H c_\theta^3 - 3 v_s \lambda_{H\Phi} c_\theta^2 s_\theta +3 v \lambda_{H\Phi} c_\theta s_\theta^2- 6 v_s \lambda_\Phi s_\theta^3 ,\\
g_{hhh'}=& 6 v \lambda_H c_\theta^2 s_\theta 
+ 6 v_s \lambda_\Phi c_\theta s_\theta^2 
+ v_s \lambda_{H\Phi} (c_\theta^3 - 2 c_\theta s_\theta^2)
+ v \lambda_{H\Phi} (s_\theta^3 -2 c_\theta^2 s_\theta),\\
g_{hh'h'}=& 6 v \lambda_H c_\theta s_\theta^2
- 6 v_s \lambda_\Phi c_\theta^2 s_\theta 
+ v \lambda_{H\Phi} (c_\theta^3 - 2 c_\theta s_\theta^2)
+ v_s \lambda_{H\Phi} (-s_\theta^3 + 2 c_\theta^2 s_\theta),\\
g_{h'h'h'}=& 6 v \lambda_H s_\theta^3 + 3 v_s \lambda_{H\Phi} c_\theta s_\theta^2 +3 v \lambda_{H\Phi} c_\theta^2 s_\theta + 6 v_s \lambda_\Phi c_\theta^3.
\end{align}
Thus the Wilson coefficient $C_q^S$ is obtained as
\begin{align}
 C_q^S & = - 2\sum_{j,k=1,2} \frac{g_{ff h_j}g_{\chi\chi h_j h_k} g_{\chi \chi h_k}}{m_q m_{h_j}^2}  \frac{1}{(4\pi)^2} B_0(m_\chi^2, m_{h_k}^2, m_\chi^2) \nonumber \\
& +\sum_{j=1,2} \frac{g_{ff h_j} g_{VVh_j}}{m_qm_{h_j}^2} \frac{g_D^2}{4} \frac{1}{(4\pi)^2} 
\left[ (-m_V^2 + 4 m_\chi^2) \frac{ \partial}{\partial m_V^2}B_0(m_\chi^2, m_V^2, m_\chi^2) \right. \nonumber \\
& \left. \phantom{ (-m_V^2 + 4 m_\chi^2) \frac{ \partial}{\partial m_V^2}B_0(m_\chi^2, m_V^2, m_\chi^2)} 
-B_0(m_\chi^2, m_V^2, m_\chi^2) + B_0(0, m_V^2, m_V^2)\right] \nonumber \\
& - \sum_{j,k,l=1,2} \frac{g_{ff h_j} g_{\chi\chi h_j} g_{\chi \chi h_k} g_{h_j h_k h_l}}{m_q m_{h_j}^2} \frac{1}{(4\pi)^2} C_0(m_\chi^2,m_\chi^2, 0,m_k^2,m_\chi^2,m_l^2)\, .
\end{align}

On the other hand, it is relatively complicated to calculate the diagram (e) with the QCD NLO contribution.
It is given in Ref.~\cite{Ishiwata:2018sdi} as
\begin{align}
 i \mathcal{M}_e =  i\frac{\alpha_s}{\pi} G^a_{\mu\nu} G^{a \mu\nu} \sum_{j\leq k} \frac{ y_{tt h_j}y_{tth_k}}{m_\chi^4} \frac{g_{\chi\chi h_j} g_{\chi\chi h_k}}{4} J^{jk}_\mathrm{box} \, ,\label{eq:amplitude_e}
\end{align}
\begin{equation}
 J^{jk}_\mathrm{box} \equiv \frac{(2 - \delta^{jk})}{8 (4\pi)^2} \int _0 ^\infty \dd t  \frac{t I(t)}{(t+x_j)(t+x_k)} \left(1-\sqrt{\frac{t+4}{t}}\right) \, ,
\end{equation}
\begin{equation}
 I(t) \equiv \frac{t-2x_t}{t(t+4x_t)} + \frac{2 x_t ( t+x_t)}{t^2(t+4x_t)\beta} \log \left(\frac{2x_t + t(1+\beta)}{2x_t + t(1-\beta)}\right) \, ,
\end{equation}
with $\beta = \sqrt{1+4x_t /t}$, $x_j \equiv m_{h_j}^2 / m_\chi^2$, $x_t \equiv m_t^2/m_\chi^2$, and $m_t$ being the top quark mass.
Comparing with the result in Ref.~\cite{Ishiwata:2018sdi}, note the presence of the factor $4$ in the denominator in Eq.~\eqref{eq:amplitude_e} because $\chi$ is now the complex scalar instead of the real one.
Then the Wilson coefficient is obtained as
\begin{align}
 C_g ^S = \sum_{j\leq k} \frac{ y_{tt h_j}y_{tth_k}}{m_\chi^4} \frac{g_{\chi\chi h_j} g_{\chi\chi h_k}}{4} J^{jk}_\mathrm{box} \,.
\end{align}
The loop integrations $B_0$ and $C_0$ are numerically evaluated by \texttt{LoopTools}~\cite{Hahn:1998yk}.

\bibliographystyle{ptephy} 
\bibliography{pNG-SU2}

\end{document}